\documentclass[11pt]{article}\pdfoutput=1
\usepackage{cite}
\usepackage{amsmath,amsfonts,amssymb}
\usepackage[small,bf,hang]{caption}
\usepackage{slashed}
\usepackage{amsmath}
\usepackage{mathabx,amsmath}
\usepackage{latexsym,epsfig}
\usepackage{arydshln}

\def\hybrid{
        \topmargin -20pt
        \oddsidemargin 0pt
        \headheight 0pt \headsep 0pt
        \textwidth 6.25in 
        \textheight 9.5in 
        \marginparwidth .875in
        \parskip 5pt plus 1pt \jot = 1.5ex}

\hybrid

 \csname
@addtoreset\endcsname{equation}{section}

\def\moth{\mathsurround=0pt}
\newdimen\zo \zo=0pt

\def\tick{\leaders\hrule height 0.5ex depth 0pt \hskip 0.5pt}
\def\upboxfill{$\moth \setbox\zo\hbox{\tick}%
  \hskip 3pt\hbox to 0pt{$\tick$\hss}\hrulefill \hbox to 7.5pt{$\tick$\hss}$}

\def\dtick{\leaders\hrule height .34pt depth 0.5ex \hskip 0.5pt}
\def\downboxfill{$\moth \setbox\zo\hbox{\dtick}%
  \hskip 2pt\hbox to 0pt{$\dtick$\hss}\hrulefill \hbox to 2pt{$\dtick$\hss}$}

\def\bec{\begin{center}}
\def\ec{\end{center}}

\def\m{\mu}

\def\be{\begin{equation}}
\def\ee{\end{equation}}
\def\bea{\begin{eqnarray}}
\def\eea{\end{eqnarray}}
\def\ba{\begin{array}}
\def\ea{\end{array}}

\begin{document}

\begin{titlepage}
\rightline{}
\rightline{May 2018}
\begin{center}
\vskip 2cm
{\Large \bf{Leibniz-Chern-Simons Theory \\[1.5ex]
and 
Phases of Exceptional Field Theory}
}\\
\vskip 2.2cm

{\large\bf {Olaf Hohm${\,}^1$ and Henning Samtleben${\,}^2$}}
\vskip 1.6cm
{\it ${}^1$
Simons Center for Geometry and Physics, Stony Brook University,\\
Stony Brook, NY 11794-3636, USA}\\
ohohm@scgp.stonybrook.edu
\vskip .2cm

{\it ${}^2$ Univ Lyon, Ens de Lyon, Univ Claude Bernard, CNRS,\\
Laboratoire de Physique, F-69342 Lyon, France} \\
{henning.samtleben@ens-lyon.fr}

\end{center}

\bigskip\bigskip
\begin{center} 
\textbf{Abstract}

\end{center} 
\begin{quote}

We discuss a generalization of Chern-Simons theory in three dimensions 
based on Leibniz (or Loday) algebras, which are generalizations of Lie algebras. 
Special cases of such theories appear in gauged supergravity, where 
the Leibniz algebra is defined in terms of the global (Lie) symmetry algebra of the ungauged limit and 
an embedding tensor. 
We show that the Leibniz algebra 
of generalized diffeomorphisms in exceptional field theory can similarly be obtained from a  
Lie algebra that describes the enhanced symmetry of an `ungauged phase' of the theory. 
Moreover, we show that a `topological phase' of 
${\rm E}_{8(8)}$ exceptional field theory 
can be interpreted 
as a Chern-Simons theory for an 
algebra unifying the three-dimensional 
Poincar\'e algebra 
and the Leibniz algebra of ${\rm E}_{8(8)}$
generalized diffeomorphisms.

\end{quote} 
\vfill
\setcounter{footnote}{0}
\end{titlepage}

\tableofcontents


\section{Introduction}

Chern-Simons actions define topological field theories in three dimensions (3D) \cite{Deser:1981wh} 
and arise in numerous contexts, for instance: as part 
of string theory and supergravity compactified to (or constructed in) 3D \cite{deWit:2004yr}; 
as a powerful framework for knot theory \cite{Witten:1988hf}; and 
as effective field theories for the quantum Hall effect (see \cite{Witten:2015aoa} for a review). 
Moreover, pure gravity 
and supergravity in 3D have an interpretation as Chern-Simons 
theories \cite{Achucarro:1987vz,Witten:1988hc}, as have their higher-spin generalizations, 
which in turn led to new toy-models for AdS/CFT \cite{Henneaux:2010xg,Campoleoni:2010zq}.  

In general, a Lie algebra that admits an invariant quadratic form defines a gauge invariant 
Chern-Simons action for a Yang-Mills gauge field in 3D. In this paper we will 
show that there is a larger class of algebraic structures that allow for consistent Chern-Simons 
theories: the Leibniz (or Loday) algebras \cite{LODAY}. They are defined by a `product' that is not necessarily 
antisymmetric but satisfies a Jacobi-like identity. In the case that the product is antisymmetric, 
this identity coincides with the Jacobi identity and hence the algebra reduces to a Lie algebra. 
Genuine Leibniz algebras do exist, however, and define a gauge invariant Chern-Simons action, 
provided they admit a quadratic form satisfying suitable invariance conditions.  
Such algebras and their associated Chern-Simons actions have already appeared in 
the literature, notably in duality covariant formulations of gauged supergravity (in the `embedding tensor 
formalism' \cite{Nicolai:2000sc,Nicolai:2001sv}) 
and of 11-dimensional or type IIB supergravity 
(in `exceptional field theory'  \cite{Hohm:2013jma,Hohm:2014fxa,Hohm:2017wtr}). 
In this paper we will discuss Leibniz-Chern-Simons theories from a more abstract 
point of view 
that allows us, at least partially, to elevate the analogy between gauged supergravity and exceptional field theory
to a technically precise correspondence.

Exceptional field theory (ExFT) is a formulation of the spacetime actions of 11-dimensional 
or type IIB supergravity 
that is covariant under the U-duality groups E$_{d(d)}$, $d=2,\ldots, 9$. To this end, the spacetime 
is extended, in the spirit of double field theory \cite{Siegel:1993th,Hull:2009mi,Hohm:2010jy,Hohm:2010pp}, 
so that the coordinates transform covariantly under E$_{d(d)}$, 
subject to (duality covariant) section constraints. 
ExFT was developed in \cite{Hohm:2013pua,Hohm:2013vpa,Hohm:2013uia}; see \cite{deWit:1986mz,Hillmann:2009ci,Berman:2010is,Coimbra:2011ky,Berman:2012vc,Cederwall:2013oaa,Hohm:2013nja,Hohm:2015xna,Abzalov:2015ega,Musaev:2015ces,Berman:2015rcc,Baguet:2016jph,Bossard:2017aae} for previous and 
subsequent work. 
In this paper we will mainly focus on the ${\rm E}_{8(8)}$ ExFT \cite{Hohm:2014fxa}, 
whose  bosonic field content  consists
of a `dreibein' $e_{\mu}{}^{a}$, an ${\rm E}_{8(8)}$ valued metric ${\cal M}_{MN}$, $M,N=1,\ldots, 248$, 
and two gauge vectors 
$A_{\mu}{}^{M}, B_{\mu M}$. All fields depend on the 248 coordinates $Y^M$ in the adjoint of ${\rm E}_{8(8)}$,  
subject to the section constraints, and on (unconstrained) external 3D coordinates 
$x^{\mu}$. 
The theory is invariant under generalized 
external and internal diffeomorphisms of the $x^{\mu}$ and $Y^M$, respectively. 
The internal diffeomorphism symmetry, when properly  formulated, 
is governed by a Leibniz algebra rather than a Lie algebra.   
In particular, the vector fields, which act as gauge fields for the generalized internal diffeomorphisms,  
naturally combine into a Leibniz valued gauge field ${\cal A}_{\mu} \equiv (A_{\mu}{}^{M},\, B_{\mu M})$, 
and enter the action precisely in a Leibniz-Chern-Simons form  \cite{Hohm:2017wtr}.

As one of our main technical results, we exhibit the close parallel between the 
Leibniz algebra structures (and their 
Chern-Simons actions) in gauged supergravity and ExFT by showing that in both frameworks 
the Leibniz algebras can be obtained by means of the same universal construction using an 
`embedding tensor'. Specifically, in gauged supergravity the structure constants of the gauge algebra are 
defined in terms of a Lie algebra $\frak{g}$ that encodes the global symmetry of the ungauged theory,  
and an embedding tensor, which in 3D is a symmetric second rank tensor on the dual space $\frak{g}^*$.  
Typically, the embedding tensor is degenerate and not invariant under the action of $\frak{g}$, 
which implies that the resulting structure constants in general do \textit{not} define a 
Lie algebra on $\frak{g}^*$. 
They define, however, a Leibniz algebra \cite{Strobl1,Strobl2}. We will then show that there is a completely analogous 
construction in ExFT, starting from an `ungauged phase' that is invariant under significantly 
enhanced global symmetries. In contrast to the full ExFT, this symmetry is governed by a genuine 
Lie algebra: the semi-direct sum of the Lie algebra of (infinitesimal) 248-dimensional diffeomorphisms 
and the current algebra of $Y$-dependent $\frak{e}_{8(8)}$ transformations. 
The quadratic invariant of the ${\rm E}_{8(8)}$ generalized diffeomorphisms can then be taken 
as the embedding tensor, which yields precisely the expected Leibniz algebra.

As a further application of the general framework of Leibniz-Chern-Simons theories, 
we will show that a certain topological subsector of the ${\rm E}_{8(8)}$
ExFT can be interpreted as a Chern-Simons theory based on an 
enlarged Leibniz algebra. This `topological phase' consists of a (covariantized) 
3D Einstein-Hilbert term 
and topological terms for the gauge vectors.
Pure 3D gravity has an interpretation as a Chern-Simons theory based 
on the Poincar\'e or (A)dS group \cite{Achucarro:1987vz,Witten:1988hc}, and we will show here that 
there is an enlarged Leibniz algebra combining the Poincar\'e algebra with the algebra of 
generalized diffeomorphisms, with the former acting on the latter by certain `anomalous' transformations.  
We show that this algebra can again be obtained from an infinite-dimensional Lie algebra $\frak{g}$ and 
an embedding tensor on $\frak{g}^*$ that acts as the symmetric invariant of the full Leibniz algebra. 
The corresponding Chern-Simons action precisely reproduces the topological sector of the ${\rm E}_{8(8)}$ ExFT, 
and we prove that the resulting gauge transformations are equivalent to those following from \cite{Hohm:2014fxa}, 
as it must be for consistency.

 One may view this theory as a 3D Chern-Simons 
theory with an infinite-dimensional `gauge group', whose algebra structure is encoded 
in the $Y$-dependence of all fields and gauge parameters.\footnote{This is 
similar to Vasiliev's higher-spin gravity in 3D, whose higher-spin algebra is defined through the dependence 
on additional coordinates \cite{Vasiliev:1992ix}, with a Chern-Simons formulation 
for the topological sector.} 
Accordingly, the 
theory still encodes genuinely 11-dimensional dynamics (or 10-dimensional dynamics, 
depending on the solution of the section constraint) and in particular 
is invariant under 11-dimensional diffeomorphisms, 
albeit formulated for a `$3+8$ foliation'. 
While this theory is topological and hence does not describe Einstein (super-)gravity in $D=11$, 
it is part of the full ${\rm E}_{8(8)}$ ExFT that encodes the complete 11-dimensional supergravity.

Formally, this topological phase is obtained by setting 
${\cal M}_{MN}=0$ in the action and gauge transformations. Of course, this is not strictly legal in that 
${\cal M}_{MN}$ was assumed to be ${\rm E}_{8(8)}$ valued and hence invertible, 
but we will show that setting ${\cal M}_{MN}=0$ does respect all gauge symmetries. 
Thus, while the resulting theory is not expected to be a consistent truncation (in the technical sense 
that any solution of the truncated theory can be uplifted to a solution of the full theory) 
it is nonetheless `consistent' by itself in that it has as much gauge symmetry as the full theory. 
In particular, this allows us, for this subsector, to make 
the external diffeomorphism symmetry manifest, which in the conventional 
formulation acts in an intricate way and so far could only be verified by tedious  computations.

We close with some general remarks.  The topological subsector of the ${\rm E}_{8(8)}$ ExFT, 
for which we here provide a Chern-Simons interpretation, is obtained by truncating the `physical' degrees of 
freedom that in 3D are entirely encoded in ${\cal M}_{MN}$. A natural and certainly legal way to do so 
would be to set it to a constant invertible matrix, say ${\cal M}_{MN}=\delta_{MN}$. 
However, any such choice would break part of the duality symmetry, here from  ${\rm E}_{8(8)}$ to ${\rm SO}(16)$, 
while the topological theory still features the full ${\rm E}_{8(8)}$ duality. 
Thus, this theory appears to be some kind of `unbroken phase'. While we have no a priori reason to assume 
that this theory by itself has some physical role to play within string/M-theory, the fact that it exists and has 
such a natural Chern-Simons interpretation is certainly intriguing. 
Regardless of whether the topological sector does or does not make physical sense by itself, 
it is part of the full ${\rm E}_{8(8)}$ ExFT, 
and so it would be important to see whether the Leibniz algebra structure also simplifies the 
`matter couplings' including ${\cal M}_{MN}$, a question to which we hope to come back to. 
We will also show that the large Leibniz algebra can be modified to (A)dS gravity. 
Again, it would be important to investigate whether this (topological) AdS theory by itself has a physical interpretation within M-theory.

The remainder of this paper is organized as follows. In sec.~2 we discuss Leibniz algebras and their 
associated Chern-Simons theories in an `invariant' (or index-free) formulation 
that is appropriate for general applications. 
Sec.~3 is mainly a review of the Leibniz algebra underlying the internal gauge symmetries 
of the ${\rm E}_{8(8)}$ ExFT. 
In this we hope to present several results that are scattered through the literature, see 
\cite{Hohm:2014fxa,Hohm:2017wtr,Baguet:2016jph}, in a 
self-contained fashion. Then we turn in sec.~4 to one of our main constructions, 
to show that the Leibniz algebra of ${\rm E}_{8(8)}$ generalized diffeomorphisms can be interpreted 
in terms of a suitably formulated embedding tensor formalism. 
Finally, in sec.~5, we define a Leibniz algebra 
combining (an infinite-dimensional extension of) the 3D Poincar\'e algebra and the ${\rm E}_{8(8)}$ generalized 
diffeomorphisms. We prove that the resulting Chern-Simons theory is equivalent to the topological 
subsector of the ${\rm E}_{8(8)}$ ExFT. We also present a generalization that includes a cosmological 
constant. Our conclusions and outlook are in  sec.~6, while in the appendix we discuss the extension 
of the embedding tensor formalism to higher dimensions.

\section{Leibniz algebras and their Chern-Simons theories} 
In this section we give a general discussion of 3D Chern-Simons theories 
based on Leibniz algebras. In the first subsection we introduce Leibniz algebras 
and their associated Yang-Mills-like vector gauge fields. In the second subsection we 
discuss the invariance conditions on a inner product and prove that the corresponding
 Chern-Simons action  is gauge invariant.

\subsection{Leibniz algebras and their gauge fields}
A Leibniz (or Loday) algebra is a vector space $X_0$ equipped with a `product'  $\circ$ 
satisfying for any vectors $x,y,z$ the Leibniz identity 
 \be\label{LeibnizID}
  x\circ (y\circ z) \ = \ (x\circ y)\circ z + y\circ (x\circ z)\;. 
 \ee
If $x \circ y$ is antisymmetric in $x, y$, this reduces to the Jacobi identity, and 
hence the algebra reduces to a Lie algebra. In the following sections we will give examples 
of genuine Leibniz algebras and thereby go beyond Lie algebras. 

An immediate consequence of (\ref{LeibnizID}) is that the product defines 
transformations 
 \be\label{genAdjoint}
  \delta_{x}y \ = \ {\cal L}_{x}y \ \equiv \ x\circ y\;, 
 \ee
that close and hence generalize the adjoint action of a Lie algebra.  
(Here we introduced the notation ${\cal L}_x$ of (generalized) Lie derivatives that will be employed later.) 
To see that (\ref{genAdjoint}) closes we compute 
\be\label{gaugealgebra}
 \begin{split}
  [{\cal L}_x, {\cal L}_y]z \ & \equiv \  {\cal L}_x({\cal L}_y z)- {\cal L}_y({\cal L}_x z) \\
  \ &= \ x\circ (y\circ z) -y\circ (x\circ z) \\
   \ &= \ (x\circ y)\circ z \\
    \ &= \ {\cal L}_{x\circ y}z\;,  
 \end{split}
 \ee  
using the Leibniz identity (\ref{LeibnizID}) from the second to the third line. 
Note that the left-hand side of (\ref{gaugealgebra}) is manifestly antisymmetric in $x, y$, 
but the right-hand side is not. Thus, antisymmetrizing on both sides of the equation we obtain 
  \be
  [{\cal L}_x,{\cal L}_y]z  
  \  = \ {\cal L}_{[x,y]}z\;, 
 \ee
while symmetrization  on both sides yields 
 \be\label{symmistrivial}
 0 \ = \  {\cal L}_{\{x,y\}}z 
  \;, 
 \ee
where we introduced the symmetrization and antisymmetrization of the product: 
 \be\label{symantisymBracket}
 \begin{split}
  \{x,y\} \ &\equiv \ \tfrac{1}{2}(x\circ y + y\circ x)\;, \\
  [x,y] \  &\equiv  \ \tfrac{1}{2}(x\circ y - y\circ x)\;. 
 \end{split} 
 \ee 
The symmetric bracket $\{\,,\}$  measures the failure of the product to define a Lie algebra. 
Importantly, even the antisymmetric  bracket $[\,,]$ does not define a Lie algebra if $\{\,,\}$ 
is non-vanishing, for then the Jacobi identity is not satisfied. However, the resulting `Jacobiator' acts 
trivially according to (\ref{symmistrivial}).

The subspace $U$ generated by symmetrized products $\{v,w\}$ forms an ideal\footnote{
The results of \cite{Hohm:2017cey} then imply that 
this algebraic structure forms part of an L$_{\infty}$ algebra~\cite{Zwiebach:1992ie}. 
See also \cite{Lavau:2014iva,Hohm:2017pnh,Cederwall:2018aab}. 
We will leave a more detailed 
discussion of the significance of such algebras in this context for future work.} 
which according to (\ref{symmistrivial}) we will refer to as the ideal of trivial vectors.
Thus ${\cal L}_u=0\,,\; \forall u\in U$\,.
In general, the Leibniz algebra may contain further vectors outside of $U$ whose generalized 
Lie derivative (\ref{genAdjoint}) on all other vectors vanishes.
In the following it will often be convenient to represent 
this ideal as the image of a linear map ${\cal D}: X_1\rightarrow U$, 
where $X_1$ is a subspace of the symmetric tensor product $X_0 \otimes_{\rm sym} X_0$
(which typically corresponds to the space of two-form gauge fields of the theory). 
Explicitly, this corresponds to a representation of the symmetrized products as
\be\label{DREL}
  \{x,y\} \ = \  \tfrac{1}{2}{\cal D}(x\bullet y)\;, 
 \ee
where $\bullet$ denotes a bilinear symmetric pairing $X_0\otimes_{\rm sym}  X_0\rightarrow X_1$. This bilinear 
map is defined by (\ref{DREL}) only up to contributions in the kernel of ${\cal D}$, which has consequences 
for the tensor hierarchies (or L$_{\infty}$ algebras) to be discussed momentarily, but it turns out that 
the related subtleties are immaterial for the 3D constructions in this paper.

After this introductory discussion, our goal is now 
to develop generalizations of Yang-Mills gauge theories for Leibniz algebras. 
In the same way that one introduces for gauge groups of Lie type one-forms taking values in the adjoint 
representation, we now introduce one-forms $A=A_{\mu}{\rm d}x^{\mu}$ taking values in 
the Leibniz algebra, of which we think as the representation space of the generalized adjoint 
action (\ref{genAdjoint}). As in Yang-Mills theory we define a gauge transformation 
w.r.t.~to a Leibniz-algebra valued gauge parameters $\lambda$: 
 \be\label{gengauge}
  \delta_{\lambda}A_{\mu} \ = \ D_{\mu}\lambda  \ \equiv \ \partial_{\mu}\lambda \ - \ A_{\mu}\circ \lambda\;. 
 \ee 
In contrast to conventional Yang-Mills theory, these transformations as such are not quite 
consistent, because they do not close by themselves. An explicit computation using (\ref{DREL}) 
shows 
 \be
 \begin{split}
 [\delta_{\lambda_1},\delta_{\lambda_2}]A_{\mu} \ = \  D_{\mu}[\lambda_2,\lambda_1] +{\cal D}(\lambda_{[1} \bullet {D}_{\mu}\lambda_{2]})\;. 
 \end{split}
 \ee 
The first term on the right-hand side takes the form of $\delta_{12}A_{\mu}$, with 
$\lambda_{12}=[\lambda_2,\lambda_1]$, but the second term spoils closure. 
This suggests to postulate a new gauge symmetry with one-form parameter $\lambda_{\mu}$ (living in the 
space $X_1$ in which $x\bullet y$ takes values): 
\be\label{FullDeltaA}
 \delta_{\lambda}A_{\mu} \ = \ D_{\mu}\lambda \  -  \   
 {\cal D}\lambda_{\mu}\;, 
\ee
for then we have closure according to  
$[\delta_{\lambda_1},\delta_{\lambda_2}]A_{\mu} = D_{\mu}\lambda_{12}-{\cal D}\lambda_{12\mu}$, where 
 \be
   \lambda_{12} \ = \ [\lambda_2,\lambda_1]\;, \qquad \lambda_{12\mu} \ = \ 
   \lambda_{[2} \bullet {D}_{\mu}\lambda_{1]}\;. 
 \ee
The one-form gauge symmetry is also needed in order for exact parameters 
 $\lambda={\cal D}a$ to yield trivial transformations. Indeed, from (\ref{FullDeltaA}) it then follows that 
 $\delta A_{\mu}=0$ for $\lambda_{\mu}=D_{\mu}a$. More precisely, here we have to assume 
 that the space in which $\lambda_{\mu}$ lives is a representation space of the Leibniz algebra, so that there is a well-defined action of ${\cal L}$ and hence a notion of covariant derivative, and that ${\cal D}$ is `covariant' 
 in that it commutes with generalized Lie derivatives. This is satisfied for all explicit examples.  
 
The new one-form gauge parameter can be associated to a new two-form gauge potential $B_{\mu\nu}$ 
taking values in the same space. Indeed, in order to define a gauge-covariant field strength such a two-form 
is needed, because the naive Yang-Mills field strength for $A_{\mu}$ in terms of the antisymmetric bracket 
$[\,, ]$ in 
(\ref{symantisymBracket}) is not gauge covariant. 
Again, the failure of covariance is ${\cal D}$ exact, and so can be fixed by setting 
 \be\label{FullCOvF}
  {\cal F}_{\mu\nu} \ = \ \partial_{\mu}A_{\nu} \ - \ 
  \partial_{\nu}A_{\mu} \ - \ [A_{\mu}, A_{\nu}]  \ + \  {\cal D}B_{\mu\nu}\;, 
 \ee
and postulating appropriate gauge transformations for $B_{\mu\nu}$. 
One may then define a gauge covariant field strength for $B_{\mu\nu}$, which in turn requires 
three-forms. This construction, which in general proceeds to higher and higher forms, 
is referred to as `tensor hierarchy'~\cite{deWit:2008ta}. 
In this paper we will focus on 3D, and it turns out that the two- and higher forms are not needed 
in order to write a gauge invariant action. Thus, we will not further develop the tensor hierarchy, 
and leave a more general discussion of tensor hierarchies for Leibniz algebras to future work.

\subsection{Invariant inner product and Chern-Simons action} 
We now turn to the construction of gauge invariant Chern-Simons actions, for which we need an 
inner product satisfying suitable invariance conditions. Thus, we assume the 
existence of a symmetric bilinear (but not necessarily non-degenerate) quadratic form, i.e., a mapping of two 
vectors $x, y$ of the Leibniz algebra to a number $\langle x, y\rangle$, satisfying $\delta_{z}\langle x,y\rangle=0$ or 
 \be\label{INVform}
  \langle z\circ x, y\rangle +\langle x,z\circ y\rangle \ = \ 0\;, 
 \ee
for arbitrary $x, y, z$.  
This property is analogous to that of invariant quadratic forms of Lie algebras. 
It turns out that we need in addition a `higher' invariance condition, corresponding to the need discussed above 
to introduce higher-form symmetries. Specifically, we need to impose 
 \be\label{innerdeg}
  \langle x,U\rangle \ = \ 0\;,  
 \ee   
for arbitrary vectors $x$ in the Leibniz algebra and 
the ideal $U$ of trivial vectors. Indeed, we can think of this condition as an invariance condition
under the `gauge transformation' $x\rightarrow x+{\cal D}a$, as in (\ref{FullDeltaA}). 
(More precisely, this would be the invariance condition of $\langle x, x\rangle$, but by polarization this 
implies the invariance of the bilinear form in general.) Note that (\ref{innerdeg}) implies that for non-trivial 
${\cal D}$ (i.e., for non-trivial $\{\,,\}$ or genuine Leibniz algebras) the bilinear form is degenerate. 

Let us next specialize to 3D and define a Chern-Simons action 
for Leibniz valued gauge vectors $A_{\mu}$. Using the inner product, we can write  
 \be\label{LeibnizChernSimons}
  S \ = \ \int {\rm d}^3x\,\varepsilon^{\mu\nu\rho}\,\big\langle  A_{\mu}\,, \,\partial_{\nu}A_{\rho}
  -\tfrac{1}{3}A_{\nu}\circ A_{\rho} \big\rangle \;, 
 \ee
where we denote by $\varepsilon^{\mu\nu\rho}$ the constant Levi-Civita symbol defining a tensor density. 
Thus, this action is manifestly invariant under 3D diffeomorphisms  and is topological.

In order to prove the gauge invariance of this action under (\ref{FullDeltaA}), 
it is convenient to first 
determine  its variation under arbitrary  $\delta A$. We compute 
 \be
  \begin{split}
   \delta_AS \ = \ \int {\rm d}^3x\,\varepsilon^{\mu\nu\rho}\,\Big(&\big\langle \delta A_{\mu}\, , 
   \, 2\,\partial_{\nu}A_{\rho}\big\rangle 
   -\tfrac{1}{3}\big\langle \delta A_{\mu}\,, \, A_{\nu}\circ A_{\rho}\big\rangle \\
   &\quad +\tfrac{2}{3}\big\langle A_{\rho},A_{\nu}\circ \delta A_{\mu}\big\rangle 
   +\tfrac{2}{3}\big\langle A_{\mu}, \{A_{\nu},\delta A_{\rho}\}\big\rangle \Big)\\
    \ = \ \int {\rm d}^3x\,\varepsilon^{\mu\nu\rho}\,\Big(&\big\langle \delta A_{\mu}\,, \,2\,\partial_{\nu}A_{\rho}
    -A_{\nu}\circ A_{\rho}\big\rangle  \ + \ 
    \tfrac{1}{3}\big\langle A_{\mu}, {\cal D}(A_{\nu}\bullet \delta A_{\rho})\big\rangle\Big)\;, 
  \end{split} 
 \ee
where we discarded a total derivative, used the invariance condition (\ref{INVform}), and (\ref{DREL}). 
We now observe that the final term in here vanishes by the `higher' invariance condition (\ref{innerdeg}). 
Moreover, for the same reason, we can add the two-form term in (\ref{FullCOvF}) to the first term 
to write the final result 
in the manifestly covariant form  
 \be\label{FinalVar}
  \delta_AS \ = \ \int {\rm d}^3x\,\varepsilon^{\mu\nu\rho}\, \big\langle \delta A_{\mu}\,, {\cal F}_{\nu\rho}\big\rangle\; . 
 \ee
At this point it is important to recall that the bilinear form in general is degenerate, so this relation does not 
imply that the field equations are ${\cal F}=0$. The field equations only imply 
that a suitable projection of the field strength vanishes. 

It is now easy to verify gauge invariance under $\delta A_{\mu}=D_{\mu}\lambda$. Inserting this transformation into 
(\ref{FinalVar}) and integrating by parts, we need to compute $D_{[\mu}{\cal F}_{\nu\rho]}$. 
In contrast to Lie algebras, this \textit{is not} zero in general, but the failure of the naive Bianchi identity 
is necessarily writable in terms of $\{\,,\}$ and thus, by  (\ref{DREL}), is  ${\cal D}$  exact. 
It then follows with (\ref{innerdeg}) that the action is invariant. Similarly, by (\ref{innerdeg}), the Chern-Simons 
action is invariant under the gauge transformations associated to the two-form, 
$\delta A_{\mu}=-{\cal D}\lambda_{\mu}$, despite the two-form not entering the Chern-Simons action.  
Summarizing, we have shown that any Leibniz algebra that admits a quadratic form satisfying the 
invariance conditions (\ref{INVform}) and (\ref{innerdeg}) defines a gauge invariant Chern-Simons action 
in 3D.

\section{Leibniz algebra of ${\rm E}_{8(8)}$ generalized diffeomorphisms}

In this section we review the gauge structure of internal generalized diffeomorphisms of the 
${\rm E}_{8(8)}$ ExFT and show that they can be interpreted as a Leibniz algebra with invariant 
quadratic form, for which the corresponding Chern-Simons action precisely yields the topological 
terms for the gauge vectors of ${\rm E}_{8(8)}$ ExFT. 

We begin by recalling a few generalities of ${\rm E}_{8(8)}$ and the associated generalized 
Lie derivatives. The Lie algebra $\frak{e}_{8(8)}$ is 248-dimensional, with generators 
$(t^M)^N{}_{K}=-f^{MN}{}_{K}$ and structure constants $f^{MN}{}_{K}$, where $M,N=1,\ldots, 248$
are adjoint indices. 
The maximal compact subgroup is SO$(16)$, under which the adjoint representation  decomposes 
as ${\bf 248}\rightarrow {\bf 120}\oplus {\bf 128}$. The invariant Cartan-Killing form   
is defined by $\eta^{MN} = \frac{1}{60} f^{MK}{}_{L} f^{NL}{}_{K}$, which we freely use to raise and lower adjoint indices.  We next need some properties of the tensor product ${\bf 248}\otimes {\bf 248}$, which 
decomposes as 
 \be\label{248squared}
  {\bf 248}\otimes {\bf 248} \ \rightarrow \ {\bf 1}\oplus  {\bf 248}\oplus {\bf 3875}\oplus {\bf 27000}
  \oplus {\bf 30380}\;.
 \ee 
It contains the adjoint ${\bf 248}$, and  the corresponding projector is given by: 
\begin{eqnarray}
\mathbb{P}^M{}_N{}^K{}_L &=&
\frac1{60}\,f^M{}_{NP}\,f^{PK}{}_L{} \\ 
&=&
\frac1{30}\, \delta_{(N}^M \delta_{L)}^K 
-\frac7{30}  (\mathbb{P}_{3875}){}^{MK}{}_{NL}
-\frac1{240}\,\eta^{MK}\eta_{NL}
+\frac1{120}\, f^{MK}{}_{P}\,f^{P}{}_{NL}{}
\;, \nonumber 
\label{pex}
\end{eqnarray}
while the projector onto the ${\bf 3875}$ reads 
  \be\label{3875proj}
   (\mathbb{P}_{3875}){}^{MK}{}_{NL} \ = \ \frac{1}{7}\,\delta^M_{(N}\, \delta^{K}_{L)}
   -\frac{1}{56}\,\eta^{MK}\,\eta_{NL}-\frac{1}{14}\,f^{P}{}_{N}{}^{(M}\, f_{PL}{}^{K)}\;. 
  \ee 
  
 We next introduce functions or fields depending on coordinates $Y^M$ living in the adjoint 
 representation, subject to the ${\rm E}_{8(8)}$ covariant `section constraints' 
   \be\label{secconstr}
  \eta^{MN}\partial_M\otimes  \partial_N \ = \ 0 \;, \quad 
  f^{MNK}\partial_N\otimes \partial_K \ = \ 0\;, \quad 
  (\mathbb{P}_{3875})_{MN}{}^{KL} \partial_K\otimes \partial_L \ = \ 0
  \;.
 \ee 
This constraint is to be interpreted in the sense that for any two fields (or gauge parameters) $A, B$ we have 
$\eta^{MN}\partial_M\partial_NA=\eta^{MN}\partial_MA\,\partial_NB=0$, and similarly for 
the other conditions in (\ref{secconstr}). These constraints are necessary in order to define consistent 
generalized Lie derivatives, to which we turn now. 
The generalized Lie 
derivative is defined with respect to two gauge parameters $\Lambda^M$, $\Sigma_M$, 
and acts on an adjoint vector $V^M$ (which may carry an intrinsic density weight $\lambda$) as 
 \be\label{usualgenLie}
  {\cal L}^{[\lambda]}_{(\Lambda,\Sigma)}V^M \ = \ \Lambda^N\partial_N V^M 
  +f^{M}{}_{NK} R^N V^K +\lambda\, \partial_N\Lambda^N V^M\;, 
 \ee
where we defined 
 \be\label{RDEF}
  R^M \ \equiv \ f^{MN}{}_{K}\,\partial_N\Lambda^K +\Sigma^M\;. 
 \ee
It is important that the gauge parameter $\Sigma_M$ is not arbitrary,  
for otherwise we could simply absorb the $\Lambda$-dependent terms in (\ref{RDEF})
into a redefinition of $\Sigma$. Rather, $\Sigma$ is `covariantly constrained' 
in the sense that it is subject to the same `sections constraints' (\ref{secconstr}) as the partial derivatives. 
Specifically, (\ref{secconstr}) holds for any two factors being partial derivatives or covariantly 
constrained, e.g., 
  \be\label{secconstr2}
  \eta^{MN}\partial_M\otimes  \Sigma_N \ = \ 0 \;, \quad 
  f^{MNK}\partial_N\otimes \Sigma_K \ = \ 0\;, \quad 
  (\mathbb{P}_{3875})_{MN}{}^{KL} \partial_K\otimes \Sigma_L \ = \ 0
  \;.
 \ee 
As a consequence of these section constraints, we have `trivial' gauge parameters, i.e., 
gauge parameters that do not generate transformations on fields.    
These include parameters of the form 
 \be\label{TrivialPara}
 \begin{split}
  &\Lambda^M \ = \ \eta^{MN}\Omega_N  \;, \\
  &\Lambda^M \ = \ (\mathbb{P}_{3875})^{MK}{}_{NL}\, \partial_K\chi^{NL}\;, \\
  & \Lambda^M \ = \ f^{MN}{}_{K}\,\Omega_{N}{}^{K}\;, \qquad 
  \Sigma_M \ = \ \partial_M\Omega_{N}{}^{N}+\partial_N\Omega_M{}^{N}\;, 
 \end{split}
 \ee
where $\Omega_M$ is covariantly constrained, and $\Omega_M{}^{N}$ is covariantly constrained in the first index.

Let us now turn to the gauge structure, which will be governed by a Leibniz algebra. 
In order to uncover this algebraic structure it is instrumental to 
group the two gauge parameters into the `doubled' object 
 \be\label{Upsilon}
  \Upsilon \ = \ \big(\Lambda^M, \Sigma_M\big)\;, 
 \ee
so that the second component is a covariantly constrained object.    
We now define the product 
 \be\label{LeibnizE8Prod}
  \Upsilon_1 \circ  \Upsilon_2 \ \equiv \ \Big(\,{\cal L}_{\Upsilon_1}^{[1]}\Lambda_2{}^M \;,\; \,
  {\cal L}_{\Upsilon_1}^{[0]}\Sigma_{2M} \ + \ \Lambda_2{}^N\partial_M R_{N}(\Upsilon_1)\,\Big)\;, 
 \ee
where the Lie derivatives act as in (\ref{usualgenLie}), with the density weights indicated in square parenthesis, 
and $R(\Upsilon)$ is defined by (\ref{RDEF}). 
The (generalized) Lie derivative terms represent the naive `covariant' action on $\Upsilon=(\Lambda,\Sigma)$, 
but the `anomalous' term containing $\partial_MR_N$ is crucial for the following. 

In order to prove that this indeed defines a Leibniz algebra it is convenient to 
use the product (\ref{LeibnizE8Prod}) to define a generalized Lie derivative on a `doubled vector' ${\cal A}$, 
with components of the same type as (\ref{Upsilon}), as  
 \be\label{extendedLIE}
  {\cal L}_{\Upsilon}{\cal A} \ \equiv \ \Upsilon \circ {\cal A}\;. 
 \ee
The Leibniz algebra relation is then equivalent to the closure condition 
 \be\label{MasterCLosure}
  \big[ {\cal L}_{\Upsilon_1}, {\cal L}_{\Upsilon_2}\big]{\cal A} \ = \ 
  {\cal L}_{[\Upsilon_1,\Upsilon_2]}{\cal A}\;, 
 \ee
where the bracket $[\,,]$ is the antisymmetrization of the Leibniz algebra (\ref{LeibnizE8Prod}), c.f.~(\ref{symantisymBracket}). The equivalence of the above closure condition to the Leibniz algebra 
relation follows as in (\ref{gaugealgebra}). 
The proof of (\ref{MasterCLosure}) proceeds by an explicit computation. 
We do not display this computation, apart from noting the useful relations 
 \be\label{RMMasterREL}
  R_M([\Upsilon_1,\Upsilon_2]) \ = \ 2\, \Lambda_{[1}{}^{N}\partial_N R_M(\Upsilon_{2]}) 
  \ + \ f_{MNK} R^N(\Upsilon_1)R^K(\Upsilon_2)\;, 
 \ee
 which is sufficient for proving closure of (\ref{usualgenLie}), and 
 \be\label{closureEXT}
  \partial_M R_N([\Upsilon_1,\Upsilon_2]) \ = \ 
  {\cal L}_{\Upsilon_1}^{[-1]}\big(\partial_MR_N(\Upsilon_2)\big)    
  - {\cal L}_{\Upsilon_2}^{[-1]}\big(\partial_MR_N(\Upsilon_1)\big)    \;, 
 \ee
which can be verified by taking the derivative of (\ref{RMMasterREL}) 
and using the Lemma  (2.13) of \cite{Hohm:2014fxa}.  
For more details we refer to Appendix A in \cite{Hohm:2017wtr}.

According to the general scheme discussed in sec.~2, the symmetrization of the Leibniz product 
(\ref{LeibnizE8Prod}) is by construction `trivial'. As a consistency check, this can 
be verified with an explicit computation: 
 \be\label{trivialE8Leibniz}
  \begin{split}
   \{\Upsilon_1 ,  \Upsilon_2 \}
     \ = \ \Big(\,&7(\mathbb{P}_{3875})^{MK}{}_{NL}\, 
   \partial_K\big(\Lambda_1^N\Lambda_2^L\big) + \tfrac{1}{8}\, \partial^M\big(\Lambda_1^N \Lambda_{2N}\big)
   +f^{MN}{}_{K}\, \Omega_{N}{}^{K}\;, \\
   &\qquad \partial_M\Omega_N{}^{N} + \partial_N\Omega_{M}{}^{N}\;\Big)\;, 
  \end{split}
 \ee   
where 
 \be
  \Omega_{M}{}^{N} \ = \ \Lambda_{(1}{}^{N}\Sigma_{2)M}
  -\tfrac{1}{2}\, f^{N}{}_{KL}\,\Lambda_{(1}{}^{K}\,\partial_M\Lambda_{2)}{}^L\;. 
 \ee
This is indeed  of the `trivial' form (\ref{TrivialPara}), in particular, $\Omega_{M}{}^{N}$ defined here 
is manifestly covariantly constrained in the first index, which is carried by either $\Sigma_M$ or $\partial_M$. 
We can further spell out the decomposition (\ref{DREL}) for the E$_{8(8)}$ Leibniz algebra, 
defining the bilinear operation 
$\Upsilon_1 \bullet \Upsilon_2$ by stripping off the derivatives in (\ref{trivialE8Leibniz}) (and multiplying 
by an overall factor of 2). The vector space $X_1$ in which $\bullet$ takes values thus decomposes 
into different subspaces, corresponding to the different terms in (\ref{trivialE8Leibniz}),
and to the two-form gauge fields in the theory, c.f.~\cite{Hohm:2014fxa}.  
Finally, the operator ${\cal D}$ acts differently on these subspaces, its action being defined
by the derivatives in (\ref{trivialE8Leibniz}) (and the inclusion map for covariantly constrained terms). 

Let us now turn to the definition of an invariant quadratic form on the Leibniz algebra. 
For doubled, Leibniz valued fields ${\cal A}=(A^M, B_M)$ it is given by 
 \be\label{invariantproduct}
  \langle  {\cal A}_1, {\cal A}_{2}\rangle \ \equiv \ \int {\rm d}^{248}Y\big(2
  A_{(1}{}^M B_{2)M} - f^{M}{}_{NK}A_{(1}{}^{N} \partial_MA_{2)}{}^{K}\big)\;. 
 \ee 
The invariance condition (\ref{INVform}) is equivalent to the statement that this integral is invariant 
under the variations (\ref{extendedLIE}), which one may verify by an explicit computation. 
In particular, both terms carry density weight one and thus vary into a total derivative that 
vanishes under the integral, up to `anomalous' contributions originating in the first term from the anomalous transformations of $B$ and in the second term from the non-covariance of partial derivatives. 
An explicit computation shows that these anomalous terms precisely cancel.  
(See Appendix A in \cite{Hohm:2017wtr} for more details.) 
Discarding total derivatives, the bilinear form can also be written as 
 \be\label{firstinvariant}
  \langle  {\cal A}_1, {\cal A}_{2}\rangle \ \equiv \ \int {\rm d}^{248}Y\big(
  A_1{}^M B_{2M}+A_2{}^M B_{1M} - f^{M}{}_{NK}A_1{}^{N} \partial_MA_2{}^{K}\big)\;, 
 \ee 
and consequently, in terms of $R_M$ defined in (\ref{RDEF}), as 
 \be
   \langle  {\cal A}_1, {\cal A}_{2}\rangle \ \equiv \ \int {\rm d}^{248}Y\big(
   A_1{}^{M}R_{M}({\cal A}_2) +A_2{}^M B_{1M}\big)\;. 
 \ee
This form is convenient in order to establish the second invariance condition (\ref{innerdeg}) 
in the form 
  \be\label{degeneracy}
  {\cal T}\quad {\rm trivial} \qquad \Rightarrow \qquad \big\langle {\cal A}\,,
  {\cal T}\big\rangle   \ = \ 0
  \quad \forall \; {\cal A}\;. 
 \ee 
This follows because for trivial ${\cal T}$ we have $R_M({\cal T})=0$, as one 
may quickly verify, and the contraction of the first component of a trivial ${\cal T}$
with a covariantly constrained $B_M$ vanishes.

Having established the Leibniz algebra relations and the existence of an invariant 
quadratic form, we can now define a Chern-Simons action for Leibniz algebra valued 
gauge vectors 
 \be\label{E8gaugevector}
  {\cal A}_{\mu} \ = \ (A_{\mu}{}^{M}, B_{\mu M})\;. 
 \ee
Their gauge transformations are given by (\ref{gengauge}) w.r.t.~an algebra valued gauge parameter 
$\Upsilon=(\Lambda^M, \Sigma_M)$.  In components these are determined with 
(\ref{LeibnizE8Prod}) to be 
\be
\begin{split}
 \delta A_{\mu}{}^{M} \ &= \ D_{\mu}\Lambda^M \;, \\
   \delta B_{\mu M} \ &= \ D_{\mu}\Sigma_{M}
   -\Lambda^N\partial_MR_{N}({\cal A}_{\mu})\;, 
\end{split}
\ee   
where here and in the following we use the covariant derivative 
 \be\label{covDER}
  D_{\mu} \ = \ \partial_{\mu} -{\cal L}_{{\cal A}_{\mu}}\;. 
 \ee
The associated field strengths ${\cal F}_{\mu\nu}=(F_{\mu\nu}{}^{M}, G_{\mu\nu M})$ 
for (\ref{E8gaugevector}) can be defined as usual through the commutator of covariant derivatives, 
 \be\label{fieldSStrengths}
  [D_{\mu}, D_{\nu}] \ = \ -{\cal L}_{(F_{\mu\nu}, G_{\mu\nu})}\;,
 \ee
up to trivial contributions.
Evaluating the Chern-Simons action (\ref{LeibnizChernSimons}) for ${\cal A}_{\mu}$ 
and the Leibniz algebra (\ref{LeibnizE8Prod}), using the invariant inner product (\ref{invariantproduct}), 
yields  
\bea
S_{\rm CS} &=&
\int {\rm d}^3x\, {\rm d}^{248}Y\, \varepsilon^{\mu\nu\rho}\,\Big( 
{F}_{\mu\nu}{}^M B_{\rho}{}_M
-f_{KL}{}^N \partial_\mu A_\nu{}^K \partial_N A_\rho{}^L
-\frac23\,f^N{}_{KL} \partial_M\partial_N A_{\mu}{}^K A_\nu{}^M A_{\rho}{}^L
\nonumber\\
&&{}
\qquad\qquad\qquad \qquad \qquad
-\frac13 \,f_{MKL} f^{KP}{}_Q f^{LR}{}_S\,A_\mu{}^M \partial_P A_\nu{}^Q \partial_R A_\rho{}^S
\Big) 
\;. 
\label{CS}
\eea
Here, $F_{\mu\nu}{}^M$ denotes the components of the
field strength defined as in (\ref{FullCOvF}) (which we may or may not 
take to include 2-forms, as these drop out upon contraction with $B_{\rho  M}$). 
We record for later use  the general variation of the action w.r.t.~$\delta A$, $\delta B$: 
 \be\label{genCSvar}
  \delta S_{\rm CS} \ = \ 
\int {\rm d}^3x\, {\rm d}^{248}Y\, \varepsilon^{\mu\nu\rho}\,\Big(\delta A_{\mu}{}^{M} \Big( G_{\nu\rho M}
+f_M{}^{N}{}_{K}\partial_N F_{\nu\rho}{}^{K}\Big) \ + \ \delta B_{\mu M} \,F_{\nu\rho}{}^{M}\Big)\;,  
 \ee
which immediately follows from (\ref{FinalVar}) and (\ref{firstinvariant}). 
The above action coincides with the topological action given for the E$_{8(8)}$ ExFT in \cite{Hohm:2014fxa}, 
and we have thus shown that that term has an interpretation as a Leibniz-Chern-Simons theory.

\section{Embedding tensor and ungauged phase}

The goal of this section is, first,  to show how the embedding tensor of gauged supergravity defines a 
Leibniz algebra in terms of the global symmetry (Lie) algebra of ungauged supergravity
and, second, to show that there is an analogous construction for E$_{8(8)}$ generalized 
diffeomorphisms. Specifically, we give 
an (infinite-dimensional) Lie algebra containing 248-dimensional diffeomorphisms 
and  E$_{8(8)}$ rotations whose coadjoint action defines, in terms of the bilinear form of the previous section, 
the Leibniz algebra of E$_{8(8)}$ generalized diffeomorphisms.

\subsection{Review of embedding tensor }

We begin by reviewing gauged supergravity in the embedding tensor 
formulation~\cite{Nicolai:2000sc,Nicolai:2001sv,deWit:2002vt}. 
The embedding tensor $\Theta_{M}{}^{\alpha}$ is a tensor under some duality group $G$, 
which is the global symmetry of the ungauged theory.  
This tensor encodes  
the subgroup of $G$ that is gauged. Specifically, one defines the `structure constants' 
 \be\label{defofX}
  X_{MN}{}^{K} \ = \ \Theta_{M}{}^{\alpha}(t_{\alpha})_N{}^{K} \ \equiv \ X_{[MN]}{}^{K} + Z^{K}{}_{MN}\;, 
 \ee
where indices $\alpha, \beta, \ldots$ label the adjoint of $G$, and indices $M, N,\ldots$ label a representation (typically thought of as 
the `fundamental' representation), 
and $(t_{\alpha})_N{}^{K}$ are the generators in this representation.  
This representation is the $G$-representation in which the vector fields $A_{\mu}{}^{M}$ of the ungauged theory transform, so that the covariant derivatives of the gauged theory can be written as 
$D_{\mu}=\partial_{\mu}-A_{\mu}{}^{M}\Theta_{M}{}^{\alpha}t_{\alpha}$. Similarly, 
all other couplings of gauged supergravity can be written in terms of the embedding tensor 
$\Theta_{M}{}^{\alpha}$. 

To identify the Leibniz algebra in this formalism, 
note that $X_{MN}{}^{K}$ in (\ref{defofX}) is not 
necessarily antisymmetric, and in the last equality we have decomposed it into its symmetric and antisymmetric part. 
Defining matrices with components $(X_M)_{N}{}^{K}=X_{MN}{}^{K}$, one now imposes the `closure constraint'  
or `quadratic constraint' for the commutator  
 \be\label{closureconstr}
  [ X_M, X_N] \ = \ -X_{MN}{}^{K} X_K\;. 
 \ee 
This defines a Leibniz algebra \cite{Strobl2}: writing for two vectors with components $V^M$, $W^M$,  
 \be\label{LeibnizComp}
  (V\circ W)^M \ \equiv \ X_{NK}{}^{M} V^N W^K\;, 
 \ee 
the closure constraint (\ref{closureconstr}) is equivalent to the Leibniz algebra relation \cite{Strobl1,Strobl2}  
 \be\label{LeibnizIDUVW}
  U\circ (V\circ W) - V\circ (U\circ W)\ = \ (U\circ V)\circ W \;. 
 \ee 
We can infer from (\ref{closureconstr}), by symmetrizing on both sides of the equation, 
 \be\label{nullCondition}
  Z^{K}{}_{MN} \,X_K \ = \ 0 \qquad \Rightarrow \qquad  Z^{K}{}_{MN}\, \Theta_{K}{}^{\alpha} \ = \ 0\;, 
 \ee
where we used the non-degeneracy of the Cartan-Killing form 
$\kappa_{\alpha\beta} \ \propto \ (t_{\alpha})_N{}^{K}(t_{\beta})_K{}^{N}$ to infer the second equation. 
In the above notation we have 
 \be\label{symmetricZ}
  \{ V, W\}^M \ = \ Z^{M}{}_{NK} V^N W^K\;. 
 \ee
 The tensor $Z^{M}{}_{NK}$ typically decomposes into \cite{deWit:2008ta}
 \bea
 Z^{M}{}_{NK} &=& {\cal D}^{M,I}\,d_{I,NK}
 \;,
 \eea
 with the index $I$ running over the space $X_1$ of two-form gauge potentials. The above decomposition (\ref{DREL})
 then corresponds to maps
 \bea
 (V\bullet W)_{I} &=& 2\,d_{I,MN}\,V^M W^N\;,\qquad
 ({\cal D} U)^M ~=~ {\cal D}^{M,I}\,U_I
 \;.
 \eea

We now specialize to 3D. In this case the fundamental $G$-representation in which vector fields are transforming
is given by the coadjoint representation. This follows because vector fields are introduced as duals to the 
Noether currents of the global symmetry group $G$ of the ungauged theory.
Expanding a local $G$ transformation as $\Lambda_M(x) \, t^M$ in terms of generators $t^M$, the Noether
currents are obtained by the corresponding variation of the Lagrangian into 
$\delta {\cal L}  =  \partial_{\mu}\Lambda_MJ^{\mu M}$.  
Defining the vector field strengths  through $F_{\m\nu}{}^M =\epsilon_{\mu\nu\rho} J^{\rho M}$, 
we finally learn that vector fields $A_{\mu}{}^{M}$ transform in the coadjoint representation of $G$.
(Of course for the finite-dimensional groups appearing in gauged supergravity, the adjoint
and coadjoint representation are typically equivalent.)
As a result, the embedding tensor takes the form $\Theta_{MN}$, with covariant derivatives 
$D_{\mu}=\partial_{\mu}-A_{\mu}{}^{M}\Theta_{MN}t^N$, for which 
(\ref{defofX}) reduces to
 \be\label{MASTER}
  X_{MN}{}^{K} \ \equiv \ \Theta_{ML} f^{LK}{}_{N} 
  \;,
 \ee
 with $Z^{K}{}_{MN} =  \Theta_{L(M} f^{LK}{}_{N)}$.
Moreover, the embedding tensor $\Theta_{MN}$ is taken to be symmetric as
it serves to define the Chern-Simons coupling of the vector fields, see (\ref{CSsugra}) below.
We can thus define the symmetric inner product
\be\label{innermetricTheta}
  \langle V, W\rangle \ \equiv  \ \Theta_{MN}V^MW^N\;.
\ee
It satisfies  the invariance condition:  
 \be
 \begin{split}
  \langle \xi\circ V, V\rangle  \ &= \ 
  \Theta_{MN}(X_{KL}{}^M\xi^K V^L) V^N \ = \ \Theta_{MN}
  \Theta_{KP} f^{PM}{}_{L} 
  \xi^K V^L V^N \\
  \ &= \ \Theta_{KP} Z^{P}{}_{NL} \xi^K V^L V^N \ = \ 0
  \;, 
 \end{split}
 \ee 
where we used (\ref{nullCondition}) in the last step. This proves that $\langle V, V\rangle$ is invariant, 
and by polarization this implies invariance of the bilinear form in general.  Similarly,
if any argument is of the form $Z^{M}{}_{NK} U^{NK}$ the inner product vanishes as 
a consequence of (\ref{nullCondition}), thereby implying the higher invariance condition (\ref{innerdeg}). 
Conversely, invariance of $\Theta_{MN}$ implies the Leibniz relations, which can be seen by contracting 
 \be
  \delta_{K}\Theta_{MN} \ \equiv \ X_{KM}{}^{L}\Theta_{LN}  + X_{KN}{}^{L}\Theta_{ML} \ = \ 0 
 \ee 
with $f^{NP}{}_{Q}$ and using the Jacobi identity in the second term. 

We can write the Chern-Simons action (\ref{LeibnizChernSimons}) 
in this formalism, using (\ref{LeibnizComp}) and (\ref{innermetricTheta}), 
 \be
   S \ = \ \int {\rm d}^3x\,\varepsilon^{\mu\nu\rho}\,\Theta_{MN} A_{\mu}{}^{M} \big(\partial_{\nu}A_{\rho}{}^{N} 
  -\tfrac{1}{3}X_{KL}{}^{N} A_{\nu}{}^{K}  A_{\rho}{}^{L}  \big) \;, 
  \label{CSsugra}
 \ee
which is the form of the Chern-Simons action in gauged supergravity. 
We have thus shown that the embedding tensor formalism for 3D gauged supergravity is a special 
case of the Leibniz-Chern-Simons theories introduced above in an `invariant' or `index-free' fashion. 
This index-free formulation is greatly advantageous for the applications in previous and subsequent  sections, where the algebras are governed by differential operators and hence 
are infinite-dimensional, so that an index notation would obscure much of the 
underlying generalized geometric structure.

\subsection{Leibniz algebras via coadjoint action of Lie algebras}  

Our next goal is to rewrite gauged supergravity relations such as (\ref{MASTER}) 
in an invariant or index-free language, 
which will be instrumental below for the infinite-dimensional generalizations based on function spaces.  
To this end we will have to 
carefully distinguish between the Lie algebra $\frak{g}$ of $G$ and its dual space $\frak{g}^*$, 
since in the infinite-dimensional context there will be no invariant metric to identify these spaces. 
We will follow the convention that adjoint vectors, i.e., elements in $\frak{g}$, 
are denoted by small latin or greek letters, 
while coadjoint vectors, i.e., elements in $\frak{g}^*$, are denoted by capital latin or greek letters. 
(Moreover, a vector or covector is typically denoted by a greek letter if it plays the role of a 
symmetry parameter.) We expand vectors and covectors w.r.t.~bases as $v=v_Mt^M$ and 
$A=A^M\tilde{t}_M$, respectively, 
where $t^M$ is a basis of $\frak{g}$, satisfying $[t^M, t^N]=f^{MN}{}_{K}\,t^K$, 
and $\tilde{t}_M$ is the dual basis. The pairing $\frak{g}\otimes \frak{g}^*\rightarrow \mathbb{R}$ 
then reads 
 \be\label{copairing} 
  A(v) \ \equiv \ A^Mv_M\;. 
 \ee
The adjoint representation is defined, for $\zeta, v \in \frak{g}$, by 
 \be
  \delta_{\zeta}v \ = \ {\rm ad}_{\zeta}v \ = \ [\zeta,v]\;. 
 \ee
We will use the notation $\delta_{\zeta}$ for general variations w.r.t.~a vector $\zeta$, but it turns out to be 
beneficial to also introduce notations such as 
${\rm ad}_{\zeta}$ if the specific representation needs to be made explicit.   
In order to define the coadjoint representation we have to specify how $\zeta\in\frak{g}$ acts 
on a coadjoint vector $A\in\frak{g}^*$ to yield a new coadjoint vector $\delta_{\zeta}A$. 
As the latter is defined by its action on an adjoint vector $ v\in\frak{g}$, we can define 
 \be\label{coadjointabstr}
  (\delta_{\zeta}A)(v) \ = \  ({\rm ad}^*_{\zeta}A)(v) \ = \ -A([\zeta,v])\;. 
 \ee
An immediate consequence is that for any pair of adjoint vectors $v, w\in \frak{g}$
 \be\label{antisymmprop}
   ({\rm ad}^*_{v}A)(w) \ = \ -  ({\rm ad}^*_{w}A)(v)\;.  
 \ee 
 The sign in the definition (\ref{coadjointabstr}) is such that the pairing (\ref{copairing}) is invariant: 
 \be\label{pairINV}
  \delta_{\zeta}(A(v)) \ \equiv \ (\delta_{\zeta}A)(v) + A(\delta_{\zeta} v) \ = \ 0\,.  
 \ee 
W.r.t.~a basis, the coadjoint action is given by $({\rm ad}_{\zeta}^*A)^M=f^{MN}{}_{K}\zeta_N A^K$.  

Let us now return to the gauged supergravity relation 
(\ref{MASTER}), defining the Leibniz algebra structure on $\mathfrak{g}^*$ in terms of the embedding tensor.
Contraction with two coadjoint vectors and one adjoint vector yields
 \be\label{thetaXContr}
  A^M B^N X_{MN}{}^{K} v_K \ = \ A^M\Theta_{ML} f^{LK}{}_{N}\, v_K B^N 
  \ = \ A^M\Theta_{ML}({\rm ad}^*_vB)^L\;. 
 \ee 
Here we recognized in the last equality the coadjoint action of $v$ on $B$. 
In order to rewrite this equation in invariant language we recall that the structure constants $X$
on the left-hand side define the Leibniz algebra according to (\ref{LeibnizComp}).
The right-hand side suggests to identify the embedding tensor $\Theta$ with a map 
 \be
  \vartheta\,: \quad \frak{g}^*\quad \rightarrow\quad \frak{g}\;,\qquad
  \vartheta(\tilde t_M)\ = \ -\Theta_{MN}\, t^N
  \;,
  \label{EmbVartheta}
 \ee
such that (\ref{thetaXContr}) takes the form
\bea
(A \circ B)(v) \ = \ -A(\vartheta({\rm ad}_v^* B)) \ = \ -
({\rm ad}_v^* B)\left(\vartheta(A)\right)
\;,
\eea
using the pairing (\ref{copairing}) between vectors and coadjoint vectors 
and the symmetry of $\Theta$ in the second equality.
Using (\ref{antisymmprop}) we may further rewrite the last term as
 \be
 \begin{split}
-\left({\rm ad}_v^* B\right)\left(\vartheta(A)\right)
 \ = \ \big({\rm ad}^*_{\vartheta(A)}\,B\big)(v)\;. 
 \end{split} 
 \ee
This shows that the Leibniz product is directly given by
 \be\label{circExpl}
  A\circ B \ \equiv \ {\rm ad}^*_{\vartheta(A)}\,B\;, 
 \ee
using the coadjoint action (\ref{coadjointabstr}) w.r.t.~$\vartheta(A)\in \frak{g}$. 
In particular, we can rewrite  the generalized Lie derivative w.r.t.~$\Lambda\in \frak{g}^*$ defined 
as in sec.~2 as 
 \be\label{generalgeneralLIE}
  \delta_{\Lambda}A \ \equiv \ {\cal L}_{\Lambda}A \ \equiv \ {\rm ad}^*_{\vartheta(\Lambda)}A\;.  
 \ee

We next observe that  the map defined in 
(\ref{EmbVartheta}) canonically induces a bilinear form on the dual space, 
 \be\label{ThetaINV}
  \Theta\;:\quad \frak{g}^*\,\otimes\ \frak{g}^*\quad \rightarrow\quad \mathbb{R}\;,
 \ee 
by the relation 
\bea
\label{vartheataDEF}
\Theta(A,B)&=& -A(\vartheta(B))
\;.
\eea
The fact that $\Theta$ is typically degenerate means that $\vartheta$ is not invertible: 
in general there is no map $\frak{g}\rightarrow \frak{g}^*$.  Put differently, if  
for all $A$ we have $A(\vartheta(B))=0$ then $\vartheta(B)=0$, 
but we cannot conclude that $B=0$.  
In terms of $\Theta$, we can now equivalently rewrite (\ref{thetaXContr}) as 
 \be\label{invLeibniz}
  (A \circ B)(v) \ = \ \Theta(A , \,{\rm ad}_{v}^*\,B)\;.   
 \ee

In the remainder of this subsection we will prove, within this invariant formulation, that the Leibniz algebra 
relations follow from the invariance of $\Theta$ under the gauge transformations defined by 
$\Theta$ itself via (\ref{invLeibniz}). 
Starting from the invariance condition on $\Theta$, i.e., that for all $A, B, \Lambda\in\frak{g}^*$ 
 \be\label{THetaINV}
  \delta_{\Lambda}\Theta(A,B)  \ \equiv \ 
  \Theta(\Lambda\circ A, B)+\Theta(A, \Lambda\circ B) \ = \ 0\;, 
 \ee
we first prove invariance of the pairing $(A\circ B)(v)$ with (\ref{invLeibniz}): 
 \be\label{LeibnizcompStep}
  \begin{split}
   \delta_{\Lambda}((A\circ B)(v)) \ &= \ 
   \Theta(\Lambda\circ A,{\rm ad}^*_{v}B) + \Theta(A, {\rm ad}^*_{v}(\Lambda\circ B)
   +{\rm ad}^*_{{\rm ad}_{\vartheta(\Lambda)}v}B)\\
   \ &= \ \Theta(\Lambda\circ A,{\rm ad}^*_{v}B)  +\Theta(A,{\rm ad}^*_{v}({\rm ad}^*_{\vartheta(\Lambda)}B) 
   +{\rm ad}^*_{[\vartheta(\Lambda),v]}B) \\
   \ &= \ \Theta(\Lambda\circ A,{\rm ad}^*_{v}B) +\Theta(A,\Lambda\circ ({\rm ad}^*_{v}B))\\
   \ &= \ 0\;. 
  \end{split}
 \ee   
Here we used, from the second to the third line, that the coadjoint action satisfies the Lie algebra relation, 
and we used the invariance (\ref{THetaINV}) in the last step. On the other hand, we can also write out 
the left-hand side of (\ref{LeibnizcompStep}) directly to obtain 
 \be\label{LeibnizSTEPPP}
   0 \ = \ ((\Lambda\circ A)\circ B)(v) + (A\circ (\Lambda\circ B))(v)
   +(A\circ B)({\rm ad}_{\vartheta(\Lambda)}v)\;. 
 \ee
The last term here can be written with (\ref{pairINV}) as 
 \be
  (A\circ B)({\rm ad}_{\vartheta(\Lambda)}v)
  \ = \ -({\rm ad}^*_{\vartheta(\Lambda)}(A\circ B))(v)
  \ = \ -(\Lambda\circ(A\circ B))(v)\;. 
 \ee
Back-substitution in (\ref{LeibnizSTEPPP}) shows that the Leibniz relations hold 
upon pairing with $v$. This holds for arbitrary $v$, which is sufficient to prove the Leibniz relations
since
\bea\label{nondegen}
\forall v\,: \;\;{A}(v) \ = \ 0 
\qquad\Longrightarrow\qquad A \ = \ 0
\;.
\label{non-deg}
\eea

Can one also prove the converse, that the Leibniz relations imply invariance of $\Theta$? 
This is possible, but only under the assumption that the Lie algebra $\frak{g}$ has trivial center. 
We first note that the Leibniz relations imply, 
as in sec.~2, that the above generalized Lie derivative acts trivially if the parameter equals 
a symmetrized bracket, c.f.~(\ref{symmistrivial}): 
 \be\label{STEPPINVVTHETTTAA}
   \forall A\,:\;\;  0 \ = \ {\cal L}_{\{\Lambda_1,\Lambda_2\}}A \ = \ {\rm ad}_{\vartheta(\{\Lambda_1,\Lambda_2\})}^*A 
  \qquad \Rightarrow \qquad 
   \forall v\,:\;\; {\rm ad}_{\vartheta(\{\Lambda_1,\Lambda_2\})}v\ = \ 0\;,  
 \ee
where the inference follows by pairing the first equation with $v\in\frak{g}$, using (\ref{pairINV}) 
and the property 
\bea\label{nondegen2}
\forall A\,: \;\;{A}(v) \ = \ 0 
\qquad\Longrightarrow\qquad v \ = \ 0
\;.
\eea
The last equation in (\ref{STEPPINVVTHETTTAA}) 
means $[\vartheta(\{\Lambda_1,\Lambda_2\}),v]=0$ for all $v$, such that
the vanishing center of $\mathfrak{g}$ implies that  
 \be\label{higherinvTheta}
  \vartheta(\{\Lambda_1,\Lambda_2\}) \ =  \ 0
  \qquad \Rightarrow \qquad \Theta(A,\{\Lambda_1,\Lambda_2\}) \ = \ 0 \;,  
 \ee
where the last inference follows upon pairing with $A\in\frak{g}^*$ and using (\ref{vartheataDEF}). 
Under this assumption we can now prove invariance of $\Theta$:   
 \be\label{INVVTHETTA345}
 \begin{split}
  \delta_{\Lambda}\Theta(A,A) \ &= \ 2\,\Theta(\Lambda\circ A, A) \ = \ -2\,(\Lambda\circ A)(\vartheta(A)) \\
  \ &= \ -2\,\Theta(\Lambda, {\rm ad}^*_{\vartheta(A)}A) \ = \ -2\,\Theta(\Lambda, \{A,A\}) \ = \ 0\;, 
 \end{split}
 \ee
where we used (\ref{invLeibniz}) in the third equality and (\ref{higherinvTheta}) in the last equality.

\subsection{E$_{8(8)}$ generalized diffeomorphisms
and the ungauged phase}

Our goal is to identify a Lie algebra from which the Leibniz algebra of E$_{8(8)}$
generalized diffeomorphisms can be derived by means of a suitable embedding tensor. 
In gauged supergravity, this Lie algebra is the global symmetry of the ungauged limit,
in which the embedding tensor is set to zero. Specifically, this limit removes 
the connection terms insides covariant derivatives, reducing them to partial derivatives, 
and also eliminates the potential and Chern-Simons term. 
We will now try to identify a similar `phase' of ExFT by setting to zero the analogous terms of 
the E$_{8(8)}$ ExFT action, which yields 
 \be\label{ungaugedExFT}
  S \ = \ \int {\rm d}^3x\, {\rm d}^{248}Ye\left(R 
    +  \tfrac{1}{240}\,\partial^{\mu}{\cal M}^{MN}\,\partial_{\mu}{\cal M}_{MN}\right)\;. 
 \ee 
Here $R$ is the familiar 3D Einstein-Hilbert term, without any further covariantizations. 
We note that while all fields depend on $x$ and $Y$, no $Y$-derivatives $\partial_M$ have been kept. 
In a sense, the different Fourier modes of the fields have been decoupled, and we will see in a moment that 
this leads to a significant symmetry enhancement.  

This unusual looking theory is actually completely analogous to that obtained from conventional 
(super-)gravity by compactifying,   
say, on a torus but \textit{without} truncation and then taking the `decompactification limit'.  
To make this point more transparent consider the Fourier expansion of a generic field 
on a torus $T^d$,   
\be
 \phi(x,y) \ = \ \sum_{{\bf k}\in\mathbb{Z}^d}\varphi_{{\bf k}}(x) \exp\Big(i\frac{{\bf k}\cdot y}{R}\Big)\;, 
\ee
with torus coordinates $y\cong y+2\pi R$, where 
we restored the radius $R$ (that for simplicity we take to be equal for all radii). 
The covariant derivatives emerging in Kaluza-Klein on a torus then take the schematic form 
 \be
  D_{\mu} \ = \ \partial_{\mu} - A_{\mu}{}^{m}\partial_m +\cdots \qquad \Rightarrow\qquad 
  D_{\mu} \ = \ \partial_{\mu} - \frac{1}{R}\sum_{\bf k} i\, A_{\mu}\cdot {\bf k}  +  \cdots \;, 
 \ee
where $\partial_m=\frac{\partial}{\partial y^m}$ are the internal derivatives. 
We observe that the inverse radius $\frac{1}{R}$ (or, equivalently, the Kaluza-Klein 
mass scale) acts as the coupling constant of the gauging. Thus, taking 
the `decompactification' or `zero mass' limit $R\rightarrow \infty$ equals the ungauged limit, 
in which covariant derivatives reduce to partial derivatives. Similarly, it is easy to convince oneself that 
all other couplings due to gauging, such as potential terms, disappear in this limit, confirming that (\ref{ungaugedExFT}) 
reasonably plays the role of the ungauged limit.\footnote{It is often claimed that compactifying on a circle of radius $R$ and then sending $R\rightarrow \infty$ gives back the original, uncompactified theory. 
The above considerations make clear, however, that one obtains rather an `ungauged phase' such as 
(\ref{ungaugedExFT})  that is quite different from any conventional theory.}

Having identified the `ungauged phase' of the E$_{8(8)}$ ExFT, let us now inspect its 
surviving symmetries. We claim that they are given by 
 \be
 \begin{split}
  \text{local external diffeomorphisms}: \qquad &\xi^{\mu}(x,Y)\;, \\
  \text{global E$_{8(8)}$ rotations}: \qquad &\sigma_{M}(Y)\;,\\
  \text{global internal diffeomorphisms}:\qquad &\lambda^M(Y)\;.  
 \end{split}
 \ee   
Here we refer to a symmetry as `local' if its parameter may depend on the 
external $x$ coordinates and as `global' if its parameter only depends on $Y$. 
Indeed, in order to establish the parallel to gauged supergravity, we have to think of the 
$Y$-dependence as parametrizing an infinite-dimensional global symmetry (rather than a 
finite-dimensional local symmetry).

Let us now inspect these symmetry transformations in more detail.   
The $\xi^{\mu}$ act like usual 3D diffeomorphisms, which are a manifest invariance 
of (\ref{ungaugedExFT}) since there are no $\partial_M$ derivatives that could detect the $Y$-dependence 
of $\xi^{\mu}$. The global internal 
diffeomorphisms with parameter $\Lambda^M$ act on the external dreibein as in the full ExFT: 
 \be
  \delta_{\lambda} e_{\mu}{}^{a} \ = \ \lambda^N\partial_N e_{\mu}{}^{a} 
  +\partial_N\lambda^N e_{\mu}{}^{a}\;, 
 \ee 
while the dreibein is left invariant under E$_{8(8)}$ rotations w.r.t.~$\sigma$. However, 
for ${\cal M}$, or equivalently a coset representative ${\cal V}_A{}^{M}$, 
the variations look different than in the full ExFT: 
 \be\label{deltaVungauged}
  \delta_{(\lambda, \sigma)}{\cal V}_{A}{}^{M} \ = \ \lambda^N\partial_N{\cal V}_{A}{}^{M}
  +f^{MN}{}_{K}\sigma_N{\cal V}_{A}{}^{K}\;. 
 \ee 
The E$_{8(8)}$ rotation is a manifest invariance of (\ref{ungaugedExFT}), and the $\lambda$ variations  
of the action combine into a total derivative. 
It would seem to be more natural to have the generalized Lie derivative (\ref{usualgenLie})
w.r.t.~$\lambda$ acting on ${\cal V}$
(we cannot use the normal Lie derivative because of ${\cal V}$ being E$_{8(8)}$ valued), 
but this is actually equivalent 
under the parameter redefinition $\sigma_M  \rightarrow \sigma_M+ f_M{}^{N}{}_{K}\partial_N\lambda^K$.   
In contrast to the parameter $\Sigma_M$ in the full ExFT, 
here we take  $\sigma_M$ to be unconstrained, so this is a legal redefinition. Thus, in presence of unconstrained $\sigma_M$ transformations it makes no difference whether we use the 
generalized Lie derivative or the simplified form (\ref{deltaVungauged}).

We will now identify the global symmetry Lie algebra $\frak{g}$ of the above ungauged phase, 
which can be determined from the closure relations of (\ref{deltaVungauged}). 
One finds that the Lie bracket for functions $\zeta=(\lambda^M, \sigma_M)$
is given by  
 \be\label{globalllLLie}
 \begin{split}
  [\zeta_1, \zeta_2] \ = \ \big(2\,\lambda_{[1}{}^N\partial_N\lambda_{2]}{}^M\,,\;
  2\,\lambda_{[1}{}^{N}\partial_N\sigma_{2]M} + f^{KL}{}_{M}\sigma_{1K}\sigma_{2L}\big) \;. 
 \end{split}
 \ee  
Lie algebras of this form are naturally associated to any given Lie algebra $\frak{g}_0$ 
(which here is $\frak{e}_{8(8)}$) as follows. First, for an arbitrary manifold $M$, the set $\frak{L}$ of 
smooth maps $M\rightarrow \frak{g}_0$ forms an infinite-dimensional Lie algebra, 
with the natural Lie bracket obtained from $\frak{g}_0$.  
Second, the Lie algebra $\frak{D}$ 
of (infinitesimal) diffeomorphisms on $M$ acts on $\frak{L}$ and its Lie bracket as a derivation. 
We can then define the semi-direct sum $\frak{L}\ \oplus \ {\frak D}$, whose Lie bracket is (\ref{globalllLLie}). 
(What is special about (\ref{globalllLLie}) is that the `Lie algebra indices' have the same range as 
the `world indices' of $M$; in general they need not be correlated.) 
Note that the Lie algebra $\frak{g}$ has a non-trivial ideal, given by all elements of the form $(0,\sigma)$. 
Similarly, the set of elements of the form $(\lambda,0)$ forms a subalgebra that is isomorphic to 
the diffeomorphism algebra $\frak{D}$. 

Next, we investigate the adjoint and coadjoint representations of (\ref{globalllLLie}). 
The adjoint representation acts on vectors $v=(p^M,q_M)\in \frak{g}$ according to 
$\delta_{\zeta}v  =  {\rm ad}_{\zeta}v  =  [\zeta,v]$, which yields for the components 
 \be\label{adjointRepr}
  \begin{split}
   \delta_{\zeta} p^M \ &= \ \lambda^N\partial_Np^M - \partial_N\lambda^M p^N\;, \\
   \delta_{\zeta} q_M \ &= \ \lambda^N\partial_N q_M - p^N\partial_N\sigma_M + f^{KL}{}_{M}\sigma_Kq_L\;. 
  \end{split}
 \ee
A coadjoint vector in $\frak{g}^*$ can be viewed as (doubled) functions ${\cal A}\equiv(A^M, B_M)$, 
with the pairing $\frak{g}\otimes \frak{g}^*\rightarrow \mathbb{R}$ given by 
the integral\footnote{This has a 
direct precursor in Witten's treatment of the coadjoint 
representation of the Virasoro group \cite{Witten:1987ty}, where coadjoint vectors are viewed as 
quadratic differentials, and the pairing between vectors and covectors is given by the invariant integral.
Note that this characterization of $\frak{g}^*$ yields  a smaller space than the unconstrained 
definition of the dual space as the `space of all functionals of $\frak{g}$', which would include delta distributions.}  
 \be\label{Pairing}
  {\cal A}(v) \ \equiv \ \int {\rm d}Y\big(A^M q_M \ + \ B_Mp^M\big)\;, 
 \ee
where from now on we set ${\rm d}Y\equiv {\rm d}^{248}Y$. The coadjoint action 
$\delta_{\zeta}{\cal A}={\rm ad}_{\zeta}^*{\cal A}$ is determined, 
as in (\ref{pairINV}), by requiring invariance of the integral. 
One quickly verifies that under (\ref{adjointRepr}) and 
 \be\label{coadjointAction}
  \begin{split}
   \delta_{\zeta}A^M \ &= \ \lambda^N\partial_NA^M+ f^{MN}{}_{K}\sigma_N A^K+\partial_N\lambda^N A^M\;, \\
   \delta_{\zeta}B_M \ &= \ \lambda^N\partial_NB_M + \partial_M\lambda^N B_N + \partial_N\lambda^N B_M
   +A^N\partial_M \sigma_N\;, 
  \end{split}
 \ee
the expression under the integral in (\ref{Pairing}) transforms into a total derivative, thereby proving invariance.

 We will now show that the coadjoint action (\ref{coadjointAction}) on ${\cal A}$ gives rise 
 to the E$_{8(8)}$ generalized diffeomorphisms using a simple identification of $(\lambda, \sigma)$ 
 with $\Upsilon=(\Lambda, \Sigma)$. 
 Specifically, let us define a map $\vartheta:\,\frak{g}^*\rightarrow\frak{g}$ as in (\ref{EmbVartheta}) by 
 \be\label{varthetaE8}
  \vartheta(\Upsilon) \ = \ \big(\vartheta(\Upsilon)^M, \, \vartheta(\Upsilon)_M\big)
  \ = \ \big(\Lambda^M, \; f_{M}{}^{N}{}_{K}\partial_N\Lambda^K + \Sigma_M\big)\;, 
 \ee
or, using the notation for the matrix $R_M$
defined in (\ref{RDEF}), 
 \be\label{varthetaE8next}
  \vartheta(\Upsilon) \ = \ \big(\Lambda^M, \; R_M(\Lambda,\Sigma)\big)\;. 
 \ee
Using the E$_{8(8)}$ 
Leibniz algebra (\ref{LeibnizE8Prod}) written out explicitly in the following form (which uses that $\Sigma$ is covariantly constrained, c.f.~eq.~(2.15) in \cite{Hohm:2014fxa}) 
 \be\label{explE8Leibniz}
 \begin{split}
  {\Upsilon}_1\circ {\Upsilon}_2 \ = \ \big(&\Lambda_1{}^{N}\partial_N \Lambda_{2}{}^{M} + f^{MN}{}_{K} R_N({\Upsilon_1}) \,\Lambda_2{}^K
  +\partial_N\Lambda_1{}^{N} \Lambda_2{}^{M}\,,\; \\
  &\;\; \Lambda_1{}^{N}\partial_N \Sigma_{2M} +\partial_N\Lambda_1^N \Sigma_{2M}
  +\partial_M\Lambda_1^N\Sigma_{2N} +\Lambda_2{}^{N}\partial_M R_N(\Upsilon_1) \big)\;, 
 \end{split}
 \ee 
it then becomes manifest, using the form of the coadjoint action (\ref{coadjointAction}),  
that the E$_{8(8)}$ generalized Lie derivative can 
be written as in (\ref{generalgeneralLIE}), 
 \be\label{Lieinvartheta}
  {\cal L}_{\Upsilon}{\cal A} \ = \ {\rm ad}^*_{\vartheta(\Upsilon)}{\cal A}\;. 
 \ee 
This shows that $\vartheta$ as defined in (\ref{varthetaE8}) encodes 
the expected Leibniz structure.

We now reconsider the bilinear form (\ref{invariantproduct}), whose arguments are coadjoint vectors 
${\cal A}=(A^M, B_M) \in \frak{g}^*$, with the goal to 
interpret it as the embedding tensor for 
the Leibniz algebra of E$_{8(8)}$ generalized diffeomorphisms. 
To this end, we compute the embedding tensor in the bilinear form induced by
the map $\vartheta$ defined in (\ref{varthetaE8}) according to (\ref{vartheataDEF}): 
 \bea
  \Theta({\cal A}_1,{\cal A}_2)  &=&  - {\cal A}_1(\vartheta({\cal A}_2)) \ = \ 
  - \int {\rm d}Y\big(A_1{}^{M} \vartheta({\cal A}_2)_M
  +B_{1M}\,\vartheta({\cal A}_2)^M\big)
\nonumber\\
 &=& - \int {\rm d}Y\big(
  A_1{}^M B_{2M}+A_2{}^M B_{1M} - f^{M}{}_{NK}A_1{}^{N} \partial_MA_2{}^{K}\big)\;, 
 \label{varthetaE8abs}
 \eea 
which indeed coincides with (\ref{firstinvariant}), up to an overall sign that we picked for later convenience.   
According to (\ref{invLeibniz}), the Leibniz algebra should then satisfy 
 \be\label{LeibnizE8Embedd}
  ({\cal A}_1\circ {\cal A}_2)(\zeta) \ = \ \Theta({\cal A}_1,\,{\rm ad}^*_{\zeta}\,{\cal A}_2)\;. 
 \ee
To confirm this, we evaluate the right-hand side by taking the second argument of (\ref{varthetaE8abs}) 
to be given by the coadjoint action (\ref{coadjointAction}) of ${\cal A}_2$. 
The left-hand side is evaluated with the pairing (\ref{Pairing}) and the Leibniz algebra (\ref{explE8Leibniz}). 
One finds that both sides precisely agree, proving that (\ref{varthetaE8abs}) can indeed be interpreted as the 
embedding tensor that `derives' the E$_{8(8)}$ Leibniz algebra from the Lie algebra (\ref{globalllLLie}).

Let us emphasize that the verification of (\ref{LeibnizE8Embedd}) does not require 
the use of any constraints, neither the section constraints on $\partial_M$ nor those on $B_M$. 
(More precisely, there are different ways of writing the Leibniz algebra that are only equivalent 
under the assumption of section constraints. The above verification 
without section constraints requires the form (\ref{explE8Leibniz}).) 
However, the product (\ref{explE8Leibniz}) satisfies the Leibniz relations (or, equivalently, 
defines generalized Lie derivatives that close) only provided we impose these constraints. 
Thus, from the point of view of the embedding tensor formulation, these constraints 
are needed in order to satisfy the quadratic constraints. Luckily, as proved in the previous subsection, 
the invariance of $\Theta$ under $\delta_{\Upsilon}={\cal L}_{\Upsilon}$ implies the 
Leibniz relations. As the former is easier to prove than the latter (see Appendix A in \cite{Hohm:2017wtr}), 
we have thereby simplified the discussion of the closure constraints.

Let us finally point out the following subtlety of the above construction: While the embedding tensor 
(\ref{varthetaE8abs}) is gauge invariant under the transformations defined by the Leibniz algebra, 
the map $\vartheta$ given in (\ref{varthetaE8}) 
is \textit{not} gauge invariant in the sense that 
 \be\label{OmegaE8}
  \Omega({\cal A}_1, {\cal A}_2) \ \equiv \ 
  \vartheta({\rm ad}_{\vartheta({\cal A}_1)}^*{\cal A}_2) -  {\rm ad}_{\vartheta({\cal A}_1)}\vartheta({\cal A}_2)
 \ee
does not vanish. In fact, invariance of $\Theta$ does not imply invariance of $\vartheta$ 
since by (\ref{varthetaE8abs}) this only needs to hold upon pairing with another coadjoint vector, 
whose second component is assumed to be `covariantly constrained'. As a consistency check 
one may verify that (\ref{OmegaE8}) indeed does vanish after pairing with such a coadjoint vector. 
For the same reason,  for $\Theta$ given, $\vartheta$ is not uniquely determined by (\ref{varthetaE8abs}), 
because the first component $\vartheta({\cal A})^M$ can be shifted by 
terms that vanish upon contraction with a constrained $B_M$.\footnote{Note, however, 
that this degeneracy of the adjoint/coadjoint pairing does not invalidate 
the proof around eq.~(\ref{LeibnizcompStep}) that invariance of $\Theta$ implies the 
Leibniz algebra relations, because we established the latter relations upon pairing with an 
arbitrary (unconstrained) vector $v$, and the inference (\ref{nondegen}) thus is still valid. 
In contrast, (\ref{nondegen2}) no longer holds, and thus the Leibniz algebra relations 
do not conversely imply invariance of $\Theta$. \\
\textit{Note added}: after submission of this paper 
we found a more streamlined treatment in which the `global' Lie algebra is given by a coset algebra $\frak{g}/\frak{I}$, 
so that the adjoint/coadjoint pairing is non-degenerate \cite{HohmSamtleben}.}

\section{Topological phase of ${\rm E}_{8(8)}$ exceptional field theory} 

We show that the topological subsector of the  ${\rm E}_{8(8)}$ ExFT 
has an interpretation as a Chern-Simons theory based on an extended Leibniz algebra. 
In the first subsection, we construct the extended 
${\rm E}_{8(8)}$-Poincar\'e Leibniz algebra and discuss the corresponding Chern-Simons theory.  
In the second subsection we will interpret this Leibniz algebra, as above,  
via the coadjoint action of an extended Lie algebra. 
In the third subsection we prove the equivalence of the Chern-Simons gauge transformations 
and that of the original ${\rm E}_{8(8)}$ ExFT, while the last subsection briefly discusses 
the extension to the AdS case.

\subsection{${\rm E}_{8(8)}$ Poincar\'e Leibniz algebra} 

We now show that 
the above Leibniz algebra based on E$_{8(8)}$ generalized diffeomorphisms 
can be enlarged to contain an infinite-dimensional generalization of the 3D Poincar\'e algebra. 
These Poincar\'e transformations in turn act via novel anomalous terms on the 
E$_{8(8)}$ Leibniz algebra, in a way that permits 
the existence of an invariant quadratic form on the total Leibniz algebra. 
The corresponding Chern-Simons theory will then be show to reproduce exactly the topological 
subsector of the E$_{8(8)}$ ExFT as described in the introduction.

The elements of this algebra combine parameters of the 3D Poincar\'e group and 
of the E$_{8(8)}$ Leibniz algebra discussed in the previous section, all being 
functions of $248$ coordinates: 
 \be\label{Xicomponents}
  \Xi \ = \ \big(\xi^a\,,\; \lambda_a\, ; \;\Lambda^M\,, \;\Sigma_M\big)\;,  
 \ee
where $a,b=0,1,2$ are SO$(1,2)$ indices. 
The Leibniz algebra structure is defined by 
 \be
  \Xi_1\circ \Xi_2 \ \equiv \ \big(\xi_{12}^a\,,\; \lambda_{12a}\, ; \;\Lambda_{12}^M\,, \;\Sigma_{12 M}\big)\;,  
 \ee
where  
 \be\label{E8Poincare} 
  \begin{split}
   \xi_{12}^a \ &= \ 2\,\varepsilon^{abc}\,\xi_{[1 b} \, \lambda_{2]c} \ + \ 2\, {\cal L}^{[1]}_{\Lambda_{[1}}\xi_{2]}{}^a\;, \\
   \lambda_{12a} \ &= \ \varepsilon_{abc}\,\lambda_1^b\,\lambda_2^c 
    \ + \ 2\,{\cal L}_{\Lambda_{[1}}^{[0]}\lambda_{2]a}\;, \\
   \Lambda_{12}^M \ &= \ {\cal L}_{\Upsilon_1}^{[1]}\Lambda_2^M\;, \\
   \Sigma_{12M} \ &= \ {\cal L}_{\Upsilon_1}^{[0]}\Sigma_{2M} \ + \ \Lambda_2^N\partial_M R_{N}(\Upsilon_1)
   \ - \ \frac{1}{\kappa}\, \xi_{[1}{}^a\partial_M\lambda_{2]a}\;, 
  \end{split}
 \ee  
and $\kappa$ is a free parameter.  
Moreover, we use the notation $\Upsilon\equiv (\Lambda^M, \Sigma_M)$, 
and we have employed the notation ${\cal L}$ for the  E$_{8(8)}$ generalized Lie derivatives above. 
In particular, the Poincar\'e parameters, not carrying E$_{8(8)}$ indices, are scalar (densities) of specific weights.  
Note that the last term in $\Sigma_{12}$ can be thought of as a non-central extension of the 
E$_{8(8)}$ Leibniz algebra by Poincar\'e generators and takes structurally the same form as 
the `anomalous' term $\Lambda\partial R$ whose need we discussed in sec.~3; in particular, 
due to its free index being carried by a derivative, it is manifestly compatible with the constraint on $\Sigma$. 
In contrast to the term $\Lambda\partial R$, however, the coefficient of this term is a free parameter 
in that the above satisfies the Leibniz algebra relation  
 \be\label{LeibnizSigma}
  \Xi_1\circ (\Xi_2\circ \Xi_3) - \Xi_2\circ (\Xi_1\circ \Xi_3)\ = \ (\Xi_1\circ \Xi_2)\circ \Xi_3 \;, 
 \ee
for any value of $\kappa$, as we will prove momentarily.  
Thus, we could take the limit $\kappa\rightarrow \infty$ and remove this non-central extension, 
but it turns out that a suitable invariant quadratic form only exists for finite $\kappa$.

In order to verify that (\ref{E8Poincare}) indeed satisfies the Leibniz algebra relations (\ref{LeibnizSigma}) 
it is convenient to consider the adjoint action on a vector in the Leibniz algebra,  
 \be\label{scalarcomponents}
  \mathfrak{A} \ \equiv \ \big(e^a\,,\; \omega_a\, ; \;A^M\,, \; \, B_M\big)\;, 
 \ee
defined by $\delta\, {\mathfrak A} \ \equiv \ \Xi  \circ  \mathfrak{A}$, 
and then to prove that they close, with an `effective' parameter given by the Leibniz algebra itself. 
Indeed, it is easy to see, precisely as in (\ref{gaugealgebra}), that closure is equivalent to the Leibniz algebra relations  (\ref{LeibnizSigma}). 
Using (\ref{E8Poincare}), the adjoint action reads in terms of components, 
  \be\label{comptrans}
  \begin{split}
   \delta e^a \ &= \ \varepsilon^{abc}\,\xi_b\, \omega_c - \varepsilon^{abc} e_b \lambda_c 
   +{\cal L}_{\Lambda}^{[1]}e^a - {\cal L}_{A}^{[1]}\xi^a\;, \\
   \delta \omega_{a} \ &= \  \varepsilon_{abc}\lambda^b\omega^c
   + {\cal L}_{\Lambda}^{[0]}\omega_a-{\cal L}_{A}^{[0]}\lambda_a \;, \\
   \delta A^M \ &= \ {\cal L}^{[1]}_{\Upsilon}A^M\;, \\
   \delta B_M \ &= \ {\cal L}_{\Lambda}^{[0]}B_M + A^N \partial_M R_N(\Upsilon)
   -\frac{1}{2\kappa} \,\xi^a\partial_M\omega_a + \frac{1}{2\kappa}\, e^a\partial_M\lambda_a\;. 
  \end{split}
 \ee     
Most of these variations are guaranteed to close by themselve. For instance, the Poincar\'e 
transformations w.r.t.~$\lambda$ and $\xi$ acting on $e^a$ and $\omega_a$  
close by themselves, because the Poincar\'e subsector defines a Lie algebra whose adjoint action 
closes. (This subsector does not define a subalgebra, however, because it acts non-centrally  
on the E$_{8(8)}$ part, as encoded in the last two terms in the last line of (\ref{comptrans})). 
Moreover, the E$_{8(8)}$ generalized diffeomorphisms close by themselves (and acting on 
$e^a, \omega_a$ as scalar densities), by the general results reviewed in sec.~3. 
Thus, the only non-trivial check is the closure on $B_M$ 
of variations involving the non-central variations proportional 
to $\frac{1}{\kappa}$. For instance, a quick computation shows that two Lorentz transformations 
on $B_M$ close according to 
 \be
  \big[\delta_{\lambda_1},\delta_{\lambda_2}\big]B_M \ = \ -\frac{1}{2\kappa}\, e^a\partial_M\lambda_{12 a}\;, 
 \ee
with $\lambda_{12}$ given by the algebra (\ref{E8Poincare}). 
The closure relations for the remaining parameters follow similarly, thereby completing the proof 
of (\ref{LeibnizSigma}). 
Let us also note that the trivial parameters of the above transformations are unmodified compared 
to the pure E$_{8(8)}$ case (\ref{TrivialPara}), because the modifications by  Poincar\'e parameters 
are antisymmetric. In particular, the symmetrization of the Leibniz product (\ref{E8Poincare}) 
vanishes in the first two (i.e.~Poincar\'e) components, and reduces in the E$_{8(8)}$ components 
to (\ref{trivialE8Leibniz}).

After having constructed the Leibniz algebra (\ref{E8Poincare}), our next task is to 
construct a symmetric bilinear form that is invariant in the sense of sec.~2. 
We start from the following ansatz generalizing the invariant form 
(\ref{invariantproduct}) of the pure E$_{8(8)}$ Leibniz structure: 
 \be\label{fullINV}
  \langle\mathfrak{A}_1, \mathfrak{A}_2 \rangle \ = \ 2\int {\rm d}^{248}Y\Big(
  \, e_{(1}{}^{a}\, \omega_{2)a} \ + \
  2\,\kappa\, A_{(1}{}^{M} B_{2)M} \ - \ \kappa\,f^{K}{}_{MN} A_{(1}{}^M \partial_K A_{2)}{}^N\,\Big)\;. 
 \ee
The first term added here is the symmetric invariant of the 3D Poincar\'e algebra 
(which was used by Witten to show that pure 3D gravity without cosmological constant
has an interpretation as a Chern-Simons theory of the 3D Poincar\'e group \cite{Witten:1988hc}). 
The second and third term, which we here multiplied by an overall factor $\kappa$, 
equal the bilinear form (\ref{invariantproduct}) and are hence invariant under pure E$_{8(8)}$
generalized diffeomorphisms. Thus, it remains to verify that the additional variations of the Poincar\'e
invariant due to the E$_{8(8)}$ diffeomorphisms in the first two lines of (\ref{comptrans}) 
are cancelled by the new, non-central variations of $B_M$. 
We compute with (\ref{comptrans}) 
 \be
  \delta(e_{(1}{}^{a}\, \omega_{2)a}) \ = \ {\cal L}_{\Lambda}^{[1]}(e_{(1}{}^{a}\, \omega_{2)a})
  -\partial _N\big(\xi^aA_{(1}{}^{N}\omega_{2)a}\big) 
  + A_{(1}{}^{N}\big(\xi^a\partial_N\omega_{2)a} - e_{2)}{}^{a}\partial_N\lambda_a\big)\;. 
 \ee 
The first term is the covariant variation (of weight one), as needed for invariance under an integral. 
The second term vanishes under the integral, and the third term is precisely 
cancelled in the combination (\ref{fullINV}), due to the extra variations of $B$ 
proportional to $\frac{1}{\kappa}$, thus proving the invariance of (\ref{fullINV}). 
The `higher' invariance condition (\ref{degeneracy}) follows as for the pure E$_{8(8)}$ theory, 
since the form of `trivial' parameters is unchanged. 

We see that both for the Leibniz relations as for the existence of an invariant bilinear form 
$\kappa$ is a free parameter, but it needs to be finite 
in order to have a non-degenerate quadratic invariant. 
More precisely, $\kappa$ needs to be non-zero in order for the invariant not to vanish 
for arbitrary values of $A, B$ (or alternatively $e^a, \omega_a$); the bilinear form is actually degenerate because of (\ref{degeneracy}). 
We will see that in the final topological theory the actual value of $\kappa$, as long as it is finite, 
has no physical significance in that it can be absorbed into a rescaling of the dreibein.

We can now construct the Chern-Simons gauge theory based on the Leibniz algebra (\ref{E8Poincare}) 
with quadratic invariant (\ref{fullINV}). We thus introduce one-forms $\frak{A}_{\mu}$ in 3D taking values 
in the Leibniz algebra and postulate Yang-Mills-like gauge transformations as in (\ref{gengauge}), 
 \be
  \delta \,\mathfrak{A}_{\mu} \ = \ \partial_{\mu}\Xi \ - \ \mathfrak{A}_{\mu}\ \circ \  \Xi\;. 
 \ee
Parametrizing the gauge parameter as (\ref{Xicomponents}) and the gauge field as 
 \be\label{fullLeibnizgauge}
  \frak{A}_{\mu} \ = \ \big(e_{\mu}{}^a\,,\; \omega_{\mu a}\, ; \;A_{\mu}{}^M\,, \; \, B_{\mu M}\big)\;, 
 \ee 
the gauge transformations read in components\footnote{These component fields should not be confused 
with those in (\ref{scalarcomponents}), because here we consider algebra valued one-form fields, 
as opposed to zero-form `matter fields'.} 
 \be\label{gaugevectortrans}
  \begin{split} 
   \delta e_{\mu}{}^{a} \ &= \ D_{\mu}\xi^a  - \varepsilon^{abc} e_{\mu b} \lambda_{c} 
   +\varepsilon^{abc}\, \xi_b\, \omega_{\mu c} +  {\cal L}^{[1]}_{\Lambda}e_{\mu}{}^{a} \;,  \\
   \delta \omega_{\mu a} \ &= \ D_{\mu}\lambda_a - \varepsilon_{abc}\, \omega_{\mu}{}^{b} \lambda^c 
    + {\cal L}_{\Lambda}^{[0]}\omega_{\mu a}\;, \\
   \delta A_{\mu}{}^{M} \ &= \ D_{\mu}\Lambda^M \;, \\
   \delta B_{\mu M} \ &= \ D_{\mu}\Sigma_{M}
   -\Lambda^N\partial_MR_{N}(A_{\mu}, B_{\mu}) + \frac{1}{2\kappa}\, e_{\mu}{}^{a}\partial_{M}\lambda_a
   -\frac{1}{2\kappa}\, \xi^a\partial_M \omega_{\mu a}\;, 
  \end{split} 
 \ee 
where we introduced covariant derivatives (\ref{covDER}) w.r.t.~the internal E$_{8(8)}$ generalized 
diffeomorphisms.   
Under the latter symmetries  the above are the expected gauge transformations 
for the one-form sector of the E$_{8(8)}$ ExFT, and so are the local Lorentz transformations for 
$e_{\mu}{}^{a}$ and $\omega_{\mu a}$, but not for $B_{\mu M}$, which is related to the corresponding 
field of the E$_{8(8)}$ ExFT by a field redefinition to be discussed in the next subsection.

Evaluating the Chern-Simons action (\ref{LeibnizChernSimons}), 
using the algebra (\ref{E8Poincare}) and the invariant (\ref{fullINV}), yields
 \be\label{finalEHCSAction}
  S \ = \ \int {\rm d}^3x\,{\rm d}^{248}Y\,\big(\varepsilon^{\mu\nu\rho}
  e_{\mu}{}^a R_{\nu\rho a}  \ + \ 2\,\kappa\,{\cal L}_{\rm CS}(A,B)\big)\;, 
 \ee 
with the Chern-Simons Lagrangian in (\ref{CS}) for the gauge vectors, and the (3D version of the) generalized 
Riemann tensor,\footnote{Compared to the conventions of \cite{Hohm:2014fxa,Hohm:2013jma} we have 
redefined the spin connection by $\omega\rightarrow -\omega$. Moreover, the spin connection 
is related to its standard form by the 3D redefinition 
$\omega_{\mu}{}^{ ab}=\varepsilon^{abc} \omega_{\mu c}$.} 
 \be\label{3DRiemann}
  R_{\mu\nu a} \ = \ D_{\mu}\omega_{\nu a}-D_{\nu}\omega_{\mu a} - \varepsilon_{abc}\, \omega_{\mu}{}^{b} 
  \,\omega_{\nu}{}^{c}\;, 
 \ee
where $D_{\mu}\omega_{\nu a}=\partial_{\mu}\omega_{\nu a} -A_{\mu}{}^{M}\partial_M\omega_{\nu a}$ 
are the covariant derivatives w.r.t.~internal generalized diffeomorphisms.  
The first term in (\ref{finalEHCSAction}) is the 3D form of the Einstein-Hilbert term $e R$, 
but due to the covariant derivatives in (\ref{3DRiemann}) it depends also on the gauge 
vectors $A_{\mu}{}^{M}$. 
(The non-central extension of the Leibniz algebra is needed in order for the couplings 
to $A_{\mu}{}^{M}$ to properly combine into the gauge covariant derivative.) 
The above action coincides with the topological sector of the E$_{8(8)}$ ExFT action obtained by truncating 
the `scalar' fields ${\cal M}_{MN}$, except for a field redefinition of $B_{\mu M}$, 
to which we turn below.

\subsection{E$_{8(8)}$ Poincar\'e Leibniz algebra via coadjoint action}

We now ask whether there is a similar construction to  that in sec.~4, where we showed 
that the Leibniz algebra of  E$_{8(8)}$ generalized diffeomorphisms can be obtained, as in 
gauged supergravity, from a genuine Lie algebra and a choice of embedding tensor. 
Is there a further extension of that (infinite-dimensional) Lie algebra so that 
the full E$_{8(8)}$ Poincar\'e Leibniz algebra is obtained in the same fashion? 
The answer is affirmative, as we will now show. 

The Lie algebra is defined for functions $\zeta=(\rho^a, \tau_a, \lambda^M, \sigma_M)$,
with Lie brackets 
 \be\label{BIGLIEEE}
 \begin{split}
  [\zeta_1,\zeta_2] \ = \ \Big(&2\,\varepsilon^{abc} \rho_{[1b}\,\tau_{2]c} + 
  2\,\partial_N\big(\lambda_{[1}{}^{N}\rho_{2]}{}^a\big)\,, \\
  &\;\;\varepsilon_{abc} \,\tau_{1}^b\,\tau_{2}^c + 2\,\lambda_{[1}{}^{N}\partial_N\tau_{2]a}\,, \\
  &\;\;2\,\lambda_{[1}{}^{N}\partial_N\lambda_{2]}{}^{M}\,, \\
  &\;\; 2\,\lambda_{[1}{}^{N}\partial_N\sigma_{2]M} + f^{KL}{}_{M} \sigma_{1K}\sigma_{2L}
  +2\,\alpha\, \rho_{[1}{}^{a}\partial_M\tau_{2]a}\Big)\;. 
 \end{split} 
 \ee
The parameter $\alpha$ in the last line is a free parameter, not constrained by the Jacobi identities. 
Note also the density term in the first line (for which we could also have introduced a free parameter 
that we fixed here to the final value). 
The adjoint action on $a=(n^a, m_a, p^M, q_M)$ is given by $\delta_{\zeta}a=[\zeta, a]$ 
and reads in components 
 \be
  \begin{split}
   \delta_{\zeta}n^a \ &= \ \varepsilon^{abc} \rho_b \,m_c -\varepsilon^{abc}\,n_b\,\tau_c 
   +\partial_N(\lambda^N n^a) - \partial_N(p^N\rho^a)\;, \\
   \delta_{\zeta} m_a \ &= \ \varepsilon_{abc}\,\tau^b\, m^c + \lambda^N\partial_Nm_a
   -p^N\partial_N\tau_a\;, \\
   \delta_{\zeta}p^M \ &= \ \lambda^N\partial_Np^M - p^N\partial_N\lambda^M\;, \\
   \delta_{\zeta}q_M \ &= \ \lambda^N\partial_N q_M - p^N\partial_N\sigma_M + f^{KL}{}_{M}\sigma_K q_L
   +\alpha\,\rho^a\partial_M m_a -\alpha\, n^a\partial_M\tau_a\;.   
  \end{split}
 \ee
The coadjoint representation on $\frak{A}=(e^a, \omega_a, A^M, B_M)\in\frak{g}^*$ is 
determined by demanding invariance of the pairing 
 \be
  \frak{A}(a) \ \equiv \ \int {\rm d}Y\Big(e^a m_a + \omega_{a} n^a + A^M q_M + B_M p^M\Big)\;. 
 \ee
One finds 
 \be
  \begin{split}
   \delta_{\zeta}e^a \ &= \ \varepsilon^{abc} \tau_b\, e_c + \varepsilon^{abc} \rho_b\, \omega_c
   +\partial_N(\lambda^N e^a) + \alpha\, \partial_M(A^M \rho^a) \\
   \delta_{\zeta}\omega_a \ &= \  \varepsilon_{abc}\, \tau^b \omega^c +\lambda^N\partial_N\omega_a 
   +\alpha\, A^M\partial_M\tau_a\;, \\
   \delta_{\zeta}A^M \ &= \ \partial_N(\lambda^N A^M) + f^{MN}{}_{K} \sigma_N A^K\;,  \\
   \delta_{\zeta} B_M \ &= \ \partial_N(\lambda^N B_M) + \partial_M\lambda^N B_N + A^N\partial_M\sigma_N
    - \rho^{a}\partial_M \omega_a+e^a\partial_M \tau_a\;. 
  \end{split}
 \ee 
Note that the anomalous Poincar\'e variations in the last line are not multiplied by $\alpha$, i.e., 
they survive even if we send $\alpha\rightarrow 0$ to remove the analogous term in 
the last line of (\ref{BIGLIEEE}). Thus, in this sense, this structure is an inevitable consequence 
of the coupling to the Poincar\'e algebra. 

Next, we define the map 
$\vartheta: \frak{g}^*\rightarrow \frak{g}$ by 
 \be
  \vartheta(\Xi) \ = \ \vartheta(\xi^a, \lambda_a, \Lambda^M, \Sigma_M) \ = \ 
  (\xi^a, \lambda_a, \Lambda^M, R_M(\Lambda, \Sigma))\;, 
 \ee
so that by comparing with (\ref{comptrans}) we confirm, upon setting $\alpha=-1$, 
 \be
  {\cal L}_{\Xi}\frak{A} \ = \ \Xi\circ \frak{A} \ = \ {\rm ad}_{\vartheta(\Xi)}^*\frak{A}\;, 
 \ee
up to the rescaling 
 \be\label{rescaling}
  e^a\ \rightarrow \ \frac{1}{2\kappa} e^a\;, \qquad \xi^{a} \ \rightarrow \ \frac{1}{2\kappa} \xi^a\;. 
 \ee 
We can now reconstruct the quadratic invariant as in (\ref{varthetaE8abs}): 
 \be
  \Theta(\frak{A}_1, \frak{A}_2) \ = \ -\frak{A}_1(\vartheta(\frak{A}_2))\;, 
 \ee
which reproduces (\ref{fullINV}), upon the rescaling (\ref{rescaling}), and up to the same global 
sign as in (\ref{varthetaE8abs}).

\subsection{Equivalence of gauge transformations} 

We will now prove equivalence of the gauge transformations following from the Chern-Simons formulation 
to those of the E$_{8(8)}$ ExFT constructed in \cite{Hohm:2014fxa}, and in particular give the required 
field redefinition of $B_{\mu M}$. We saw already that the Yang-Mills transformations of the Chern-Simons 
formulation give rise to the expected form of internal E$_{8(8)}$ generalized diffeomorphisms. 
Next, we turn to the external diffeomorphisms, which in the formulation of \cite{Hohm:2014fxa} 
are parametrized by $\xi^{\mu}(x,Y)$, and show that the local translations $\xi^{a}$ 
can be matched with these transformations. Note that the 
manifest diffeomorphism invariance of the Chern-Simons theory 
here only implies invariance under the $Y$-independent transformations with $\xi^{\mu}=\xi^{\mu}(x)$, 
and thus we have to identify the remaining diffeomorphisms (that in some sense mix $x$ and $Y$) 
among the infinite-dimensional Yang-Mills gauge transformations.

We begin by performing the following field-dependent redefinition of gauge parameters that introduces 
the vector parameter $\xi^{\mu}=\xi^{\mu}(x,Y)$: 
 \be
  \xi^a \ = \ \xi^{\nu} e_{\nu}{}^{a}\;, \qquad \lambda_a \ \rightarrow \ \lambda_a + \xi^{\nu}\omega_{\nu a}\;. 
 \ee
Note that at this stage we do not perform an analogous parameter redefinition of $\Lambda$, $\Sigma$. 
We then find from (\ref{gaugevectortrans}) the transformations w.r.t.~to the new parameter $\xi^{\mu}$: 
 \be\label{intermtrans}
  \begin{split}
   \delta e_{\mu}{}^{a} \ &= \ \xi^{\nu}D_{\nu} e_{\mu}{}^{a} +D_{\mu}\xi^{\nu} e_{\nu}{}^{a}
   -\xi^{\nu} T_{\nu\mu}{}^a\,, \\
   \delta \omega_{\mu a} \ &= \ \xi^{\nu}D_{\nu} \omega_{\mu a} +D_{\mu}\xi^{\nu} \omega_{\nu a} 
   -\xi^{\nu}R_{\nu\mu a}\;, \\
   \delta A_{\mu}{}^{M} \ &= \ 0\;, \\
   \delta B_{\mu M} \ &= \ \frac{1}{2\kappa}\, e_{\mu}{}^a\partial_M\lambda_a + \frac{1}{\kappa}\,
   \xi^{\nu}e_{[\mu}{}^{a}\partial_M\omega_{\nu] a} +\frac{1}{2\kappa}\, 
   e_{\mu}{}^{a} \partial_M\xi^{\nu} \omega_{\nu a}\;, 
  \end{split}
 \ee  
where $R_{\mu\nu a}$ is the (generalized) 3D Riemann tensor (\ref{3DRiemann}), and we introduced the  
generalized torsion tensor 
 \be\label{genTorsion}
  T_{\mu\nu}{}^{a} \ = \ D_{\mu}e_{\nu}{}^{a}-D_{\nu}e_{\mu}{}^{a}
  -\varepsilon^{abc} \omega_{\mu b} e_{\nu c} + \varepsilon^{abc} \omega_{\nu b} e_{\mu c}\;. 
 \ee
In order to compare with the transformations of the full ExFT, we have 
to add an equations-of-motion symmetry.  
A general such symmetry takes the form 
$\delta \mathfrak{A}_{\mu}=\Omega_{\mu\nu}\epsilon^{\nu\rho\sigma} \mathfrak{F}_{\rho\sigma}$, with 
$\Omega_{\mu\nu}$ antisymmetric. Choosing $\Omega_{\mu\nu} \ \propto \ \epsilon_{\mu\nu\rho}\xi^{\rho}$
we infer that the following provides a trivial on-shell symmetry:  
 \be\label{onshellVar}
  \delta\, \mathfrak{A}_{\mu} \ = \ \xi^{\nu}\mathfrak{F}_{\nu\mu}\;, 
 \ee
where the field strength, discussed in sec.~2, takes the form 
 \be
  \mathfrak{F}_{\mu\nu} \ = \ 2\,\partial_{[\mu}\mathfrak{A}_{\nu]}  \ - \  
  \mathfrak{A}_{[\mu}\circ \mathfrak{A}_{\nu]}+\cdots\;, 
 \ee
up to trivial 2-forms that are irrelevant in the action.  
For the vielbein and spin connection components these curvatures are given by (\ref{3DRiemann}) 
and (\ref{genTorsion}), respectively, while the curvatures $F_{\mu\nu}{}^{M}$, ${\cal G}_{\mu\nu M}$ 
for ${\cal A}_{\mu}=(A_{\mu}{}^{M}, B_{\mu M})$ 
are as discussed in sec.~3, c.f.~(\ref{fieldSStrengths}), except for the following modification: 
 \be\label{fieldstrengthREL}
  {\cal G}_{\mu\nu M} \ = \ G_{\mu\nu M} + \frac{1}{\kappa}\, e_{[\mu}{}^{a}\partial_M \omega_{\nu]a}\;, 
 \ee
due to the non-central extension of the algebra. 
Adding (\ref{onshellVar}) to (\ref{intermtrans}) we obtain equivalent gauge transformations, 
in which the field strength terms  in $\delta e$ and $\delta \omega$ are cancelled, 
while field strength terms are added to $\delta A$, 
$\delta B$: 
 \be\label{Finalcovtrans}
  \begin{split}
   \delta_{\xi}e_{\mu}{}^{a} \ &= \ \xi^{\nu}D_{\nu} e_{\mu}{}^{a} +D_{\mu}\xi^{\nu} e_{\nu}{}^{a}
  \,, \\
   \delta \omega_{\mu a} \ &= \ \xi^{\nu}D_{\nu} \omega_{\mu a} +D_{\mu}\xi^{\nu} \omega_{\nu a} 
   \;, \\
   \delta A_{\mu}{}^{M} \ &= \ \xi^{\nu} F_{\nu\mu}{}^{M}\;, \\
   \delta B_{\mu M} \ &= \ \xi^{\nu}{\cal G}_{\nu\mu M} + \frac{1}{2\kappa}\,
   e_{\mu}{}^a\partial_M\lambda_a + \frac{1}{\kappa}\,
   \xi^{\nu}e_{[\mu}{}^{a}\partial_M\omega_{\nu] a} +\frac{1}{2\kappa}\, 
   e_{\mu}{}^{a} \partial_M\xi^{\nu} \omega_{\nu a}\;. 
  \end{split}
 \ee  
Writing the gauge transformation of $B_{\mu M}$ in terms of the field strength $G_{\mu\nu}$ 
via (\ref{fieldstrengthREL}), some terms cancel, and we get 
 \be\label{intermediatedeltaB}
  \delta B_{\mu M} \ = \ \xi^{\nu}{G}_{\nu\mu M} + \frac{1}{2\kappa}\, e_{\mu}{}^a\partial_M\lambda_a 
   +\frac{1}{2\kappa}\, e_{\mu}{}^{a} \partial_M\xi^{\nu} \omega_{\nu a}\;. 
 \ee

Our goal is now to find a field redefinition so that the gauge transformations of $B_{\mu M}$ can 
be matched with those of the original E$_{8(8)}$ ExFT in \cite{Hohm:2014fxa}. 
In particular, in the latter formulation $B_{\mu M}$ is inert under local Lorentz transformations, 
while in (\ref{intermediatedeltaB}) it transforms under $\lambda_a$. 
This suggests to define a new field as 
 \be\label{Bredef}
  \bar{B}_{\mu M} \ \equiv \ B_{\mu M}  + \frac{1}{4\kappa}\,  e_{\mu}{}^{a} \varepsilon_{abc} \, \omega_M{}^{bc}\;, \qquad
  \omega_M{}^{ab} \ \equiv \  e^{\mu[a}\partial_M e_{\mu}{}^{ b]}\;,  
 \ee 
because $\omega_M$ has an anomalous Lorentz transformation $\Delta_{\lambda}\omega_{M}{}^{ab}=\varepsilon^{abc}\partial_M\lambda_c$, as can be verified with (\ref{gaugevectortrans}), 
which precisely cancels 
the $e\partial_M\lambda$ term in (\ref{intermediatedeltaB}). 
 $\bar{B}$ is then Lorentz invariant, as in the conventional 
ExFT formulation.  Performing the redefinition (\ref{Bredef}) in the action (\ref{finalEHCSAction}) 
one obtains 
\be\label{finalEHCSAction22}
  S \ = \ \int {\rm d}^3x\,{\rm d}^{248}Y\,\big(e \widehat{R}  \ + \ 2\,\kappa\,{\cal L}_{\rm CS}(A,\bar{B})\big)\;, 
 \ee 
where we defined an `improved' Riemann tensor so that  
 \be
   e\widehat{R} \ = \ \varepsilon^{\mu\nu\rho} e_{\mu}{}^{a} \widehat{R}^{}_{\nu\rho\,a}
   \ = \ eR + ee^{a\mu}e^{b\nu} F_{\mu\nu}{}^{M} \omega_{Mab}\;. 
 \ee 
This is the form of the covariantized Einstein-Hilbert term for generic ExFTs, 
where a term proportional to $F_{\mu\nu}$ is added in order to guarantee 
local Lorentz invariance. The novelty of the 3D case is that this term is not needed 
but can be absorbed into a redefinition of $B_{\mu M}$, as done by (the inverse of) (\ref{Bredef}), 
in which case the non-invariance of the Einstein-Hilbert term is compensated by a non-trivial 
Lorentz transformation of $B_{\mu M}$.

Having identified the field redefinition that matches the actions of the original and the Chern-Simons 
formulation, as well as matching the local Lorentz transformations, we prove in the remainder of this section 
that also the external generalized diffeomorphisms w.r.t.~$\xi^{\mu}$ are equivalent, 
as it should be for consistency. 
To this end we have to compute the transformation of the redefined $B_{\mu M}$ in (\ref{Bredef}) 
under (\ref{Finalcovtrans}), for which in turn we need the anomalous diffeomorphism transformation 
of $\omega_M{}^{ab}$. To this end, we compute 
\be
 \Delta_{\xi}(\partial_M e_{\mu}{}^{b}) \ = \ \partial_M\xi^\lambda D_{\lambda} e_{\mu}{}^{b}
 +D_{\mu}(\partial_M\xi^{\lambda}) e_{\lambda}{}^{b} - \xi^{\lambda} \partial_M\partial_NA_{\lambda}{}^{N} 
 e_{\mu}{}^{b}\;,   
\ee
where $\Delta_{\xi}$ denotes the non-covariant part of the diffeomorphism transformation (the difference 
between the full variation and the `covariant' terms 
that take the same form as standard infinitesimal diffeomorphisms, but with $\partial_{\mu}$ replaced by 
$D_{\mu}$). The above relation can be verified by a direct computation. 
From this we derive 
 \be\label{DeltaomegaM}
  \Delta_{\xi}\omega_{M}{}^{ab} \ = \ \partial_M\xi^{\lambda} e^{\mu[a} D_{\lambda} e_{\mu}{}^{b]}
  +D_{\mu}(\partial_M\xi^{\lambda}) e^{\mu[a} e_{\lambda}{}^{b]}\;. 
 \ee
When using (\ref{intermediatedeltaB}) in order to compute the transformation of (\ref{Bredef}) 
we may use the explicit form the spin 
connection, because in  \cite{Hohm:2014fxa,Hohm:2013jma} we employed a second order formalism 
that treats $\omega$ as determined by its own field equations, $T_{\mu\nu}{}^{a}=0$. 
It reads
 \be\label{explicitSpin}
  \omega_{\mu a} \ = \ \epsilon^{\nu\rho\sigma}\big(e_{\nu a} e_{\mu b}-\tfrac{1}{2} e_{\mu a} e_{\nu b}\big)
  D_{\rho} e_{\sigma}{}^{b}\;. 
 \ee
Moreover, we have to write the field strength in (\ref{intermediatedeltaB}) in terms of $\bar{B}$: 
 \be
  G_{\mu\nu M}(B) \ = \ G_{\mu\nu M}(\bar{B}) - \frac{1}{2\kappa}\, D_{[\mu}
  \big(e_{\nu]}{}^{a} \varepsilon_{abc}\, \omega_{M}{}^{bc}\big)\;, 
 \ee
so that one obtains for the variation of (\ref{Bredef}) 
 \be
 \begin{split}
  \delta_{\xi}\bar{B}_{\mu M} \ = \ &\xi^{\nu} G_{\nu\mu M}(\bar{B}) 
  -\frac{1}{2\kappa}\, \xi^{\nu}D_{[\nu}\big(e_{\mu]}{}^{a} \varepsilon_{abc}\, \omega_{M}{}^{bc}\big) \\
  &+\frac{1}{4\kappa}\, \xi^{\nu}D_{\nu}\big(e_{\mu}{}^{a} \varepsilon_{abc} \, \omega_M{}^{bc}\big)
  +\frac{1}{4\kappa}\, D_{\mu}\xi^{\nu}\big(e_{\nu}{}^{a} \varepsilon_{abc} \, \omega_M{}^{bc}\big) 
  +(\partial_M\xi \text{ terms}) \\
  \ = \ & \xi^{\nu} G_{\nu\mu}(\bar{B})  
  +D_{\mu}\Big(\frac{1}{4\kappa}\,\xi^{\nu}e_{\nu}{}^{a} \varepsilon_{abc}\, \omega_{M}{}^{bc}\Big)
  +(\partial_M\xi \text{ terms}) \;, 
 \end{split} 
 \ee
where we give the $\partial_M\xi$ terms momentarily.   
The terms in the second line are the covariant terms from the variation of the terms in (\ref{Bredef})
proportional to $\frac{1}{\kappa}$.  
We observe that terms combined into a total $D_{\mu}$ derivative, which can be eliminated by 
the parameter redefinition
 \be
  \bar{\Sigma}_{M} \ = \ \Sigma_M +\frac{1}{4\kappa}\,\xi^{\nu}e_{\nu}{}^{a} \varepsilon_{abc}\, \omega_{M}{}^{bc}\;.  
 \ee 
The total $\Sigma$ and $\xi$ transformations are now given by 
 \be
 \begin{split}
  \delta_{\xi,\bar{\Sigma}}\bar{B}_{\mu M} \ = \ &D_{\mu}\bar{\Sigma}_{M} + \xi^{\nu} G_{\nu\mu M} \\
  &+ \frac{1}{2\kappa}\, 
  e_{\mu}{}^{a}\Big(\partial_M\xi^{\nu}\omega_{\nu a} +\frac{1}{2} \varepsilon_{abc} \,\partial_M\xi^{\nu}
  e^{\rho b}D_{\nu}e_{\rho}{}^{c} +\frac{1}{2}\varepsilon_{abc} \, D_{\rho}(\partial_M\xi^{\nu}) e^{\rho b} 
  e_{\nu}{}^{c}\Big)\;, 
 \end{split}
 \ee
where we restored the $\partial_M\xi$ terms, using (\ref{DeltaomegaM}). Inserting (\ref{explicitSpin}) one 
finds after some manipulations,  
using a Schouten identity in the form $0=\partial_M\xi^{[\nu} \, \varepsilon^{\lambda\rho\sigma]}$, 
the following form of the external diffeomorphisms: 
 \be\label{finalDeltaB}
 \begin{split}
  \delta_{\xi}\bar{B}_{\mu M} \ &= \  \xi^{\nu} G_{\nu\mu M}
  +\frac{1}{4\kappa}\,\epsilon_{\mu\nu\sigma} D^{\nu}(\partial_M\xi^{\sigma})
  +\frac{1}{4\kappa}\,\partial_M\xi^{\nu} \epsilon_{\mu}{}^{\rho\sigma} D_{\rho} g_{\sigma \nu}\\
  \ &= \  \xi^{\nu} G_{\nu\mu M}
  +\frac{1}{4\kappa}\,\epsilon_{\mu\nu\lambda} \,g^{\lambda\rho} D^{\nu}\big(g_{\rho\sigma}\partial_M\xi^{\sigma}\big)\;. 
 \end{split} 
 \ee
The last form is precisely the gauge transformation of $B_{\mu M}$ in the original formulation (upon 
truncating the `matter' fields ${\cal M}$), 
see eq.~(3.24) in \cite{Hohm:2014fxa}. 
More precisely, in the full ExFT an on-shell modification of the gauge transformations is needed, 
in which the  field strengths terms in $\delta_{\xi}A$ and $\delta_{\xi}B$ are replaced by 
their on-shell dual `matter currents'. For the topological sector considered here we may perform another 
equations-of-motion symmetry (\ref{onshellVar}), but now only for the sector of gauge 
vectors $(A, B)$, in order to remove the field strengths terms $F_{\mu\nu}$ and $G_{\mu\nu}$,\footnote{Perhaps more simply, it is straightforward to verify with (\ref{genCSvar}) that the pure field strength terms 
in $\delta_{\xi}A_{\mu}{}^{M}$ and $\delta_{\xi}B_{\mu M}$ are a separate invariance of the Chern-Simons action 
and can hence be dropped in the formulas for external diffeomorphisms.} 
so that $\delta_{\xi}A_{\mu}{}^{M}=0$, and $\delta_{\xi}B_{\mu M}$ reduces to the 
second term in (\ref{finalDeltaB}), which agrees,  upon 
truncating the matter fields ${\cal M}$, with eq.~(3.40) in \cite{Hohm:2014fxa}. 
This completes our discussion of the proof that the gauge transformations of the topological sector of the 
E$_{8(8)}$ ExFT can be interpreted as Yang-Mills gauge transformations based on the 
Leibniz-Poincar\'e algebra (\ref{E8Poincare}).

 \subsection{Generalization to AdS gravity}

We have seen that the topological sector of E$_{8(8)}$ ExFT has a Chern-Simons interpretation, 
reproducing in particular the 3D Einstein-Hilbert term without cosmological constant. 
It is natural to ask whether there is an extension to include a non-vanishing cosmological constant, 
as is the case for pure 3D gravity, where the Poincar\'e group is simply replaced by the (A)dS groups,  
SO$(2,2)$ or SO$(1,3)$, respectively. 
Moreover, both in gauged supergravity and ExFT there is a potential, so that generic compactifications 
indeed give rise to a non-vanishing cosmological constant, thereby suggesting that 
a reformulation with a 3+8 (and eventually 3+248) split may naturally involve an external (A)dS$_3$ space. 

We will now show that there is an extension of the Leibniz algebra (\ref{E8Poincare}) 
to a de Sitter-Leibniz algebra, whose Chern-Simons action leads to a cosmological constant. 
We denote the cosmological constant by $v=-\frac{1}{\ell^2}$, with (A)dS radius $\ell$, (and we can think of it as the ground state value of the  potential 
in a complete theory, 
$V_0=v$). The Leibniz algebra is defined in terms of functions $\Xi = (\xi^a,\, \lambda_a\, ; \,\Lambda^M\,, 
\Sigma_M)$ by 
 \be
  \Xi_1\circ \Xi_2 \ \equiv \ \big(\xi_{12}^a\,,\; \lambda_{12a}\, ; \;\Lambda_{12}^M\,, \;\Sigma_{12 M}\big)\;,  
 \ee
where 
 \be\label{LeibnizdeSitter} 
  \begin{split}
   \xi_{12}^a \ &= \ 2\,\varepsilon^{abc}\,\xi_{[1 b} \, \lambda_{2]c} +2\,{\cal L}^{[1]}_{\Lambda_{[1}}\xi_{2]}^a\;, \\
   \lambda_{12a} \ &= \ \varepsilon_{abc}\,\lambda_1^b\,\lambda_2^c 
   -\frac{1}{\ell^2} \, \varepsilon_{abc}\,\xi_{1}^b\,\xi_2^c
   + 2\,{\cal L}_{\Lambda_{[1}}^{[0]}\lambda_{2]a}\;, \\
   \Lambda_{12}^M \ &= \ {\cal L}_{\Upsilon_1}^{[1]}\Lambda_2^M\;, \\
   \Sigma_{12M} \ &= \ {\cal L}_{\Upsilon_1}^{[0]}\Sigma_{2M} + \Lambda_2^N\partial_M R_{N}(\Upsilon_1)
   -\frac{1}{\kappa}\, \xi_{[1}^a\partial_M\lambda^{}_{2]a}\;. 
  \end{split}
 \ee  
The AdS length scale $\ell$ only appears in the second line, as a modification of the Lorentz sub-algebra, 
as is the case for the conventional (A)dS algebra. We can now compute the adjoint action on an algebra valued field
$\mathfrak{A} \equiv (e^a, \omega_a,  A^M,  B_M)$ 
to find the transformations 
 \be
  \begin{split}
   \delta e^a \ &= \ \varepsilon^{abc}\xi_b \omega_c - \varepsilon^{abc} e_b \lambda_c 
   +{\cal L}_{\Lambda}^{[1]}e^a - {\cal L}_{A}^{[1]}\xi^a\;, \\
   \delta \omega_{a} \ &= \  \varepsilon_{abc}\lambda^b\omega^c -\frac{1}{\ell^2}\,\varepsilon_{abc}\,\xi^b e^c
   + {\cal L}_{\Lambda}^{[0]}\omega_a-{\cal L}_{A}^{[0]}\lambda_a \;, \\
   \delta A^M \ &= \ {\cal L}^{[1]}_{\Upsilon}A^M\;, \\
   \delta B_M \ &= \ {\cal L}_{\Lambda}^{[0]}B_M + A^N \partial_M R_N(\Upsilon)
   -\frac{1}{2 \kappa} \,\xi^a\partial_M\omega_a + \frac{1}{2\kappa}\, e^a\partial_M\lambda_a\;. 
  \end{split}
 \ee     
The new term proportional to the cosmological constant  
in $\delta \omega_a$ is the only change in the transformation rules. 
The Leibniz algebra relations are thus automatically satisfied, because their equivalent 
closure conditions hold precisely as for the pure (A)dS Lie algebra. 

We next have to ask whether there is still in invariant quadratic form. 
It turns out that the invariant (\ref{fullINV}) for the Poincar\'e-Leibniz algebra continues to be invariant, 
because under the new term in the variation proportional to $v$ we have 
 \be
  \delta_{v}(e_{(1}{}^{a} \omega_{2)a}) \ = \ v\,\varepsilon_{abc}\,\xi^b \, e_{(1}{}^{a} e_{2)}{}^{c} \ = \ 0\;, 
 \ee
as a consequence of the symmetrization. Thus, we can define a Chern-Simons action 
based on (\ref{LeibnizdeSitter}), using the same invariant (\ref{fullINV}), to obtain 
  \be\label{EHdSCSAction}
  S \ = \ \int {\rm d}^3x\,{\rm d}^{248}Y\,\big(eR \ -  \ 2\,e\,v \ + \ 2\,\kappa\,{\cal L}_{\rm CS}(A,B)\big)\;.  
 \ee 
Thus, the only modification is the addition of a cosmological constant term proportional to $v$.

Let us point out a peculiar difference of the above construction to 
pure AdS$_3$ gravity and its supersymmetric and higher-spin generalizations. 
In the latter case the algebras always factorize; for instance,  
for pure gravity we have SO$(2,2)   \cong {\rm SL}(2,\mathbb{R})\times {\rm SL}(2,\mathbb{R})$, 
and the super- and higher-spin groups factorize  similarly. 
As a consequence, there is a second invariant of the Lie algebra \cite{Witten:1988hc}, which reads 
  \be\label{secondInv}
   v e_{(1}{}^{a} e_{2)a} +\omega_{(1}{}^{a}\omega_{2)a}\;, 
  \ee
which is non-degenerate for $v\neq 0$. 
Due to the splitting of the gauge groups, the Chern-Simons action is then really the 
sum of two SL$(2,\mathbb{R})$ Chern-Simons terms with arbitrary relative coefficients. 
There is no analogue for the E$_{8(8)}$ ExFT, however, because 
the second invariant cannot be consistently extended. In fact, in ExFT, $e$ carries (density) weight one 
and $\omega$ weight zero, so that in (\ref{secondInv}) both terms fail to have a total weight of one, 
as would be necessary for invariance under an integral as in (\ref{fullINV}).\footnote{There is another 3D
Chern-Simons-type theory with an infinite-dimensional extension of the AdS algebra that does not 
factorize \cite{Arvanitakis:2015sgs}. This is based on the algebra  of volume preserving 
diffeomorphisms on $S^3$ \cite{Bandos:2008jv}, which is a genuine Lie algebra that, however, does not have 
an invariant quadratic form. The Chern-Simons term constructed in \cite{Bandos:2008jv} can 
be interpreted as a Leibniz-Chern-Simons theory, with the embedding tensor being the invariant quadratic form 
on the dual space of one-forms 
given by $\Theta(\omega,\eta)=\int_{S^3}\omega\wedge {\rm d}\eta$.
We hope to elaborate on this connection in more detail 
in future work and thank an anonymous referee for inquiring about this.\\
\textit{Note added:} In the meantime, the preprint \cite{Hohm:2018git} appeared in which this connection is developed.}

We have shown that the Chern-Simons formulation can be extended so as to include 
a cosmological constant term proportional to $v$. In particular, for $v<0$ the theory admits 
AdS$_3$ solutions, and so one can investigate it as a (toy-)model for the AdS/CFT correspondence. 
As a first step it would be important to determine  the asymptotic symmetries. 
While for pure 3D gravity they are given by (two copies of) the Virasoro algebra, 
with the Brown-Henneaux central charge $c=\frac{3\ell}{2G}$, already for supergravity and 
higher-spin theories  the asymptotic symmetries are no longer governed by Lie algebras, 
but rather by so-called $W$-algebras (although the value of the central charge remains unchanged) 
\cite{Henneaux:2010xg,Campoleoni:2010zq,Henneaux:1999ib}. 
Thus, it is plausible to suspect that the same happens for the theory considered here. 
Finally we note that $W$-algebras have recently been shown to have an interpretation 
as L$_{\infty}$ algebras \cite{Blumenhagen:2017ogh}, 
as have the bulk ExFTs, and so intriguingly both the bulk and boundary 
degrees of freedom may be governed by suitable $\infty$-algebras. This may lead to a new perspective 
on holography more generally.

\section{Summary and Outlook} 

We have discussed the general construction of Chern-Simons actions in 3D based on 
Leibniz algebras and shown that they arise naturally in gauged supergravity and exceptional field theory. 
For the E$_{8(8)}$ exceptional field theory both the topological terms for the gauge vectors themselves 
and the full `topological phase' including also the dreibein and spin connection allow for such 
Chern-Simons interpretations. We have also shown that there is a universal construction of such Leibniz 
algebras that is applicable both to gauged supergravity and exceptional field theory. It starts from a 
genuine Lie algebra $\frak{g}$ (that we can view as the global symmetry of the `ungauged phase') 
and an embedding tensor, that in 3D 
is a symmetric tensor $\Theta$ on the dual space $\frak{g}^*$. Interpreting the embedding tensor 
as a map $\vartheta : \frak{g}^*\rightarrow \frak{g}$, one can define the Leibniz algebra 
in terms of the coadjoint representation as $A\circ B={\rm ad}^*_{\vartheta(A)}B$. 
The original embedding tensor $\Theta$ is invariant under this action, which in turn 
implies the Leibniz relations. 

With this construction we believe to have made a potentially significant conceptual advance 
in that it gives an answer to the question whether the generalized diffeomorphisms 
in double and exceptional field theory can be interpreted as originating from more conventional 
transformations, such as diffeomorphisms on a larger manifold, by putting additional structures. 
Obvious analogies are symplectic manifolds in which the general 
diffeomorphism group is reduced to the symplectomorphisms that leave the symplectic form  invariant. 
However, such a construction cannot give generalized diffeomorphisms, 
for starting from a Lie algebra and demanding invariance of some structure at best yields a non-trivial 
subalgebra that is still a Lie algebra, but not a genuine Leibniz algebra. (See sec.~3.1 in \cite{Hohm:2013bwa}.) 
This obstacle is circumvented in the above construction by having the `invariant structure' (the embedding tensor)  
itself define the `adjoint action' of the Leibniz algebra --- in terms of the coadjoint action of the original Lie algebra. 
Moreover, this Lie algebra is crucially not just a diffeomorphism algebra 
but rather the semi-direct sum of a diffeomorphism 
algebra and the current algebra based on the corresponding U-duality group. 
It is important to investigate this construction further, in particular  in order to see 
whether it sheds light on some of the conceptual questions of double and exceptional field theory. 

In view of the realization of the `topological phase' of the E$_{8(8)}$ exceptional field theory 
as a Leibniz-Chern-Simons theory, it remains to see whether the `matter couplings' given by 
the ${\rm E}_{8(8)}/{\rm SO}(16)$ coset degrees of freedom can be efficiently described 
in a similar language, presumably upon introducing auxiliary fields. 
Moreover, even without matter couplings, it would be interesting to see whether the topological phase 
makes physical sense by itself. Although here  we can only speculate, one may wonder whether
this Chern-Simons theory represents a protected topological 
sector of M-theory.

\subsection*{Acknowledgements}
We would like to thank Dan Butter, Franz Ciceri and Ergin Sezgin for useful discussions. 
The work of O.H. is supported by a DFG Heisenberg fellowship.

\appendix

\section*{Appendix}

\section{Embedding tensor in general dimensions} 

Although our focus in this paper is the 3D case, for completeness we discuss here how to 
define embedding tensors for generalized diffeomorphisms in arbitrary dimensions, i.e., starting from more  
general Lie algebras $\frak{g}_0$. This will illustrate from a yet different angle that the 3D case, which 
superficially seems to be rather special, fits nicely into the pattern in general dimensions. 

We start with the Lie algebra $\mathfrak{g}_0$ of the U-duality group under consideration,  
with generators $t_{\alpha}$ satisfying  $[t_{\alpha}, t_{\beta}]=f_{\alpha\beta}{}^{\gamma}t_{\gamma}$. 
Furthermore, we pick a representation space $R$ of $\mathfrak{g}_0$ in which the coordinates live, 
and write for a generic vector $v^M$, $M=1,\ldots, {\rm dim}(R)$, and for the 
representation matrices $(t_{\alpha})_M{}^{N}$. 
We can now define an infinite-dimensional extension ${\mathfrak{g}}$ of $\mathfrak{g}_0$ 
as described after (\ref{globalllLLie}). Specifically, the elements of $\frak{g}$ are functions of $Y^M$ 
denoted by $\zeta  =  (\lambda^M, \sigma^{\alpha})$,  
with Lie brackets 
 \be\label{LieBracketss}
 \begin{split}
  [\zeta_1,\zeta_2] \ = \ \big(2\,\lambda_{[1}{}^{N}\partial_N\lambda_{2]}{}^{M}\,, \;
  2\,\lambda_{[1}{}^{N}\partial_N\sigma_{2]}{}^{\alpha}
  \ + \   f_{\beta\gamma}{}^{\alpha} \sigma_1{}^{\beta}\sigma_{2}{}^{\gamma}\big) \;. 
 \end{split}  
 \ee

We now consider some important representations of this Lie algebra. First, the representation 
$R$ naturally extends to infinite-dimensional $\mathfrak{g}$ representations, whose 
elements are $R$ valued functions $v^M(Y)$, on which $\zeta\in \mathfrak{g}$ acts as 
 \be\label{bigaction}
  \delta_{\zeta}v^M \ \equiv \ 
  \rho_{\zeta}v^M \ \equiv \ \lambda^N\partial_Nv^M +  \gamma \, \partial_N\lambda^N v^M 
  - \sigma^{\alpha}(t_{\alpha})_N{}^{M}v^N\;.  
 \ee
Here $\gamma$ is an arbitrary density weight, and so we can denote this 
representation space more appropriately as $R^{[\gamma]}$. 
Using that the $(t_{\alpha})_M{}^{N}$ form a representation of the original algebra $\frak{g}_0$, 
it is straightforward to verify that (\ref{bigaction}) is indeed a representation of (\ref{LieBracketss}): 
 \be\label{representation}
  [\rho_{\zeta_1},\rho_{\zeta_2}]  \ = \ \rho_{[\zeta_1,\zeta_2]} \;. 
 \ee
We again note that the coordinate indices need not be correlated with the $R$ representation indices; 
the above would be a representation regardless, but this form is the one appearing 
in our subsequent construction.\footnote{A related question is whether in (\ref{bigaction}) 
one could employ the full Lie derivative w.r.t.~$\lambda$, which can only be written if both indices are identified. 
It turns out that this would spoil closure.}  
More generally, we can canonically define representations on any 
tensor power of $R$. In addition, we can consider the dual representation $R^*$, whose elements 
are functions $A_M$ with invariant pairing $R\otimes R^*\rightarrow \mathbb{R}$ given by 
 \be
  A(v) \ = \ \int {\rm d}Y\, v^M A_M\;, 
 \ee
where ${\rm d}Y\equiv {\rm d}^{{\rm dim}(R)}Y$.   
More precisely, if the original representation space is $R^{[\gamma]}$ the dual space $(R^{[\gamma]})^*$
consists of functions $A_M$ of intrinsic density weight $1-\gamma$, with the transformation rules 
  \be\label{starRaction}
  \delta_{\zeta}A_M \ \equiv \ \rho_{\zeta}^*A_M\ \equiv \ \lambda^N\partial_NA_M 
  + (1-\gamma)   \partial_N\lambda^N A_M
  +\sigma^{\alpha} (t_{\alpha})_{M}{}^{N}A_N\;. 
 \ee 
As usual, this definition is equivalent to 
 \be\label{copropertyyyy}
  (\rho_{\zeta}^* A)(v) \ = \ -A(\rho_{\zeta}v)\;. 
 \ee

Next, we investigate the adjoint and coadjoint representations.  
The adjoint representation acts on $a =(p^M, q^{\alpha})\in \frak{g}$ as $\delta_{\zeta}a  =  [\zeta, a]$, 
which yields in components 
 \be
  \begin{split}
   \delta_{\zeta}p^M \ &= \ \lambda^N\partial_Np^M - p^N\partial_N\lambda^M\;, \\
   \delta_{\zeta}q^{\alpha} \ &= \ \lambda^N\partial_N q^{\alpha}-p^N\partial_N\sigma^{\alpha} 
   +f_{\beta\gamma}{}^{\alpha} \sigma^{\beta} q^{\gamma}\;. 
  \end{split}
 \ee
The coadjoint representation acts on $\frak{g}^*$, whose elements are  functions 
${\cal A}  = (A_{\alpha}, B_M)$ with the pairing $\frak{g}^*\otimes \frak{g}\rightarrow \mathbb{R}$ 
defined as usual by an integral: 
 \be\label{integralPAIRING}
  {\cal A}(a) \ = \ \int {\rm d}Y\big(p^M B_M \ + \ q^{\alpha}A_{\alpha}\big) \;. 
 \ee
The coadjoint action is determined by requiring invariance of the integral  
and found to be  
 \be
 \begin{split}
  \delta_{\zeta} A_{\alpha} \ &= \ \lambda^N\partial_N A_{\alpha}+\partial_N\lambda^N A_{\alpha} 
  +f_{\alpha\beta}{}^{\gamma} \sigma^{\beta} A_{\gamma}\;, \\
  \delta_{\zeta} B_M \ &= \ \lambda^N\partial_N B_M + \partial_M\lambda^N B_N +\partial_N\lambda^N B_M
  +A_{\alpha}\,\partial_M\sigma^{\alpha}\;.  
 \end{split}
 \ee
This definition of the coadjoint representation is of course equivalent to  (\ref{coadjointabstr}). 

In order to relate to the embedding tensor formulation in arbitrary dimensions, 
we next have to use the general fact that for any representation $R$ there is a canonical map 
 \be
  \pi\,:\quad R\ \otimes \ R^* \quad \rightarrow\quad \frak{g}^*\;. 
 \ee  
(For $R$ equal to $\frak{g}$ or $\frak{g}^*$ this coincides with the coadjoint representation, 
and so this map is a natural extension of our 3D construction based on the coadjoint representation.) 
This map is defined as follows: Since its image is a coadjoint vector it naturally acts on adjoint vectors $\zeta$, 
and so we can define, for $v\in R$, $A\in R^*$, 
  \be\label{PIDEF}
  (\pi(v,A))(\zeta) \ \equiv \ (\rho^*_{\zeta}A)(v)\;. 
 \ee
To illustrate this definition we note that for a finite-dimensional Lie algebra with generators $t_{\alpha}$ 
this reads in a basis 
 \be\label{piinbasis}
   \pi(v, A)_{\alpha} \ = \ v^M(t_{\alpha})_{M}{}^{N}A_N\;. 
 \ee 
We now evaluate (\ref{PIDEF}) for our infinite-dimensional algebra and representations by computing 
for the right-hand side with (\ref{starRaction}) 
 \be
 \begin{split}
  (\rho^*_{\zeta}A)(v) 
   \ = \ \int {\rm d}Y\Big(\lambda^M\big(v^N\partial_M A_N-\partial_M\big(v^NA_N\big)
   +\gamma\, \partial_M\big(v^N A_N\big)\big)
   +\sigma^{\alpha} v^M(t_{\alpha})_{M}{}^{N}A_N \Big)
  \;, 
 \end{split}
 \ee
where we integrated by parts in order to move derivatives away from $\lambda$.  
With the pairing (\ref{integralPAIRING}) we can then read off the map $\pi$ 
from the left-hand side of (\ref{PIDEF}): 
 \be\label{explicitPI}
  \begin{split}
     \pi(v,A) \ = \ \big(v^M(t_{\alpha})_{M}{}^{N}A_N\,, \; -\partial_M v^N A_N
    +\gamma \,\partial_M\big(v^NA_N\big)\big)\;. 
  \end{split}
 \ee

We are now ready to re-interpret the embedding tensor  in these invariant terms. 
To this end we  return to the general form $\Theta_{M}{}^{\alpha}$ for the embedding tensor,  
which was employed for the gauged supergravity relation 
 \be\label{XTHetaagain}
  X_{MN}{}^{K} \ = \ \Theta_{M}{}^{\alpha}(t_{\alpha})_N{}^{K}\;, 
 \ee 
c.f.~(\ref{defofX}), and view it as a map 
 \be
  \Theta\,:\quad R\,\otimes \, \frak{g}^*\quad \rightarrow \quad \mathbb{R}\;. 
 \ee 
Contracting (\ref{XTHetaagain}) with $v, w\in R$ and $A\in R^*$ and recognizing 
the map (\ref{piinbasis}), we can write the Leibniz product on $R$ via  
 \be\label{INVEMBEDDDDDD}
  A(v\circ w) \ = \ \Theta(v, \pi(w,A))\;. 
 \ee
 
We now claim that the Leibniz algebra defined by generalized Lie derivatives in generic dimensions 
is defined through this relation upon taking the 
embedding tensor to be given, for $v\in {R}$, ${\cal A}=(A_{\alpha}, B_M)\in \frak{g}^*$, by 
 \be\label{genTheta}
  \Theta(v, {\cal A}) \ = \
   - \int {\rm d}Y\big(v^M B_M - \kappa\, (t^{\alpha})_M{}^{N} A_{\alpha}\,\partial_Nv^M\big)\;, 
 \ee
where $\kappa$ is a constant to be determined.  
Note that the last term requires an invariant bilinear form on the original Lie algebra $\frak{g}_0$ 
in order to raise the index on $t_{\alpha}$, 
which is the first time that this assumption is needed. 
Evaluating the right-hand side of (\ref{INVEMBEDDDDDD}) with (\ref{explicitPI}) we obtain 
 \be
 \begin{split}
  \Theta(v, \pi(w,A)) \ = \ \int {\rm d}Y\big( v^M\partial_Mw^NA_N
  -\gamma\, v^M\partial_M(w^NA_N)
  +\kappa (t^{\alpha})_M{}^N \partial_N v^M w^K(t_{\alpha})_K{}^{L} A_L\big)\;. 
 \end{split} 
 \ee
Integrating by parts, this can be rewritten in terms of the standard form of the generalized 
Lie derivative, 
 \be
  {\cal L}_{v}w^M \ \equiv \ v^N\partial_N w^M  + \kappa\,(t^{\alpha})_N{}^M(t_{\alpha})_L{}^{K}\,
  \partial_Kv^L\,w^N +\gamma\,\partial_Nv^N w^M\;, 
 \ee 
as follows 
 \be
  \Theta(v, \pi(w,A)) \ = \ \int {\rm d}Y \big({\cal L}_{v}w^M\big)A_M \;. 
 \ee 
Provided we choose $\lambda$ and $\kappa$, which so far are free parameters, appropriately 
(depending on the group and representation $R$), 
the right-hand side is equal to $A(v\circ w)$. This proves that the Leibniz algebra of generalized 
Lie derivatives is recovered in accordance with the embedding tensor 
construction (\ref{INVEMBEDDDDDD}).

As for the 3D case, it is illuminating to also view the embedding tensor 
as a map 
 \be
  \vartheta\,:\quad R\quad \rightarrow \quad \frak{g}\;, 
 \ee
defined, for $v\in R$, ${\cal A}\in \frak{g}^*$,  by 
 \be
  {\cal A}(\vartheta(v)) \ \equiv \ -\Theta(v,{\cal A})\;. 
 \ee
From (\ref{genTheta}) one finds that this map reads explicitly 
 \be\label{VARtheta}
  \vartheta(v) \ = \ \big(v^M\,,\;-\kappa (t^{\alpha})_{M}{}^{N}\partial_N v^M\big)\;. 
 \ee 
One of the advantages of this map is that, again, we can define  
the Leibniz product (or generalized Lie derivative) more explicitly by 
 \be\label{GENLeibnizZZ}
  v \circ w \ \equiv \ \rho_{\vartheta(v)}w\;. 
 \ee 
In order to prove that this is equivalent to (\ref{INVEMBEDDDDDD}) we act 
with a covector $A\in R^*$: 
 \be
  A(v\circ w) \ = \ A(\rho_{\vartheta(v)}w) \ = \  
  -(\rho^*_{\vartheta(v)}A)(w) \ = \ -(\pi(w,A))(\vartheta(v))
  \ = \ \Theta(v,\pi(w,A))\;, 
 \ee 
where we used (\ref{copropertyyyy}) and (\ref{PIDEF}).  
One may also quickly verify with (\ref{VARtheta}) that (\ref{GENLeibnizZZ}) yields 
the familiar formulas for generalized Lie derivatives (and thereby for the corresponding Leibniz algebras).

At this stage a cautionary remark is in order. 
In general, the embedding tensor 
map $\vartheta$ 
is not gauge invariant. 
If $\vartheta$ were gauge invariant we would have an immediate proof of the Leibniz relations as follows: 
Invariance means that, for $v, w \in R$, the following expression vanishes:  
 \be\label{OMEGA}
  \Omega(v,w) \ \equiv \ 
  \delta_{v}(\vartheta(w)) - {\rm ad}_{\vartheta(v)}\vartheta(w)  
  \ = \ \vartheta(v\circ w)  -  
  [\vartheta(v), \vartheta(w)] \;. 
 \ee 
The Leibniz relations in turn involve the combination  
 \be
 \begin{split}
  &(v_1\circ v_2)\circ w +  v_2\circ (v_1\circ w)   - v_1 \circ (v_2\circ w) \\
   & \ = \   
  \rho_{\vartheta(v_1\circ v_2)}w - [ \rho_{\vartheta(v_1)}, \rho_{\vartheta(v_2)}]w \\
  & \ =  \ \rho_{\Omega(v_1,v_2)}w\;, 
 \end{split} 
 \ee
where we used that $\rho$ forms a representation of the Lie algebra $\frak{g}$, 
and we recognized (\ref{OMEGA}) in the last step. Thus, invariance of $\vartheta$ 
or $\Omega\equiv 0$ 
implies the Leibniz relations, but the converse is not true: Due to the section constraints 
there are `trivial' parameters, so that one may have $\rho_{\Omega(v_1,v_2)}w=0$ without 
$\Omega(v_1,v_2)$ being zero. This is indeed what happens for generic ExFTs.  


\providecommand{\href}[2]{#2}\begingroup\raggedright\endgroup


\end{document}